\documentclass[smallextended]{svjour3} 
\usepackage[english]{babel}
\usepackage{graphicx}
\usepackage[utf8]{inputenc}
\usepackage{pgf,tikz,pgfplots}
\pgfplotsset{compat=1.17}
\usetikzlibrary{arrows}
\usepackage{amssymb}
\usepackage{bm}
\usepackage{csquotes}
\expandafter\let\csname equation*\endcsname\relax
\expandafter\let\csname endequation*\endcsname\relax
\usepackage{amsmath}
\usepackage{amsfonts}
\usepackage{amstext}
\usepackage{physics}
\usepackage{color}
\usepackage{hyperref}
\usepackage{enumerate}
\usepackage{graphicx}
\usepackage{float}
\usepackage{setspace} 
\usepackage{lipsum}
\usepackage{dcolumn}
\usepackage{caption}
\usepackage{subcaption}
\usepackage{epsfig}
\usepackage{epsf}
\usepackage{epstopdf}

\usepackage{enumitem}
\usepackage{xcolor}
\usepackage{tikz}
\usepackage[margin=1in]{geometry}
\usepackage{pgfplots}
\usepackage{esint}
\usepackage{scalerel}
\usepackage{stackengine}
\usepackage{braket}
\usepackage{MnSymbol}
\usepackage{color}
\usepackage{titlesec}
\usepackage{verbatim}
\usepackage{tikz}
\newcommand{\la}{\langle} 
\newcommand{\ra}{\rangle} 
\newcommand\p{\partial}
\newcommand\f{\frac}
\newcommand\e{\mathbf{e}}
\newcommand\nn{\nonumber}
\newcommand\hu{\hat{u}}
\newcommand\hv{\hat{v}}
\newcommand\hh{\hat{h}}
\newcommand\hj{\hat{j}}
\newcommand\hr{\hat{r}}
\newcommand\vf{v} %averaged velocity field
\newcommand\cf{u} %total field vector

\DeclareSymbolFont{usualmathcal}{OMS}{cmsy}{m}{n}
\DeclareSymbolFontAlphabet{\mathcal}{usualmathcal}

\newcommand{\be}{\begin{equation}}
\newcommand{\ee}{\end{equation}}
\newcommand{\bea}{\begin{eqnarray}}
\newcommand{\eea}{\end{eqnarray}}

\usepackage{color}
\definecolor{dgreen}{rgb}{0,0.7,0}

\begin{document}
\title{Hydrodynamics of the Fermi-Pasta-Ulam-Tsingou chain}

\author{Shubhadeep Chakraborty \textsuperscript{*}, Abhishek Dhar \textsuperscript{*}
}

\institute{${}^{*}$ International Centre for Theoretical Sciences, Bengaluru, India-560089}

\date{Received: date / Accepted: date}

\maketitle
\begin{abstract}
We provide a pedagogical review of the hydrodynamics of the FPUT chain. There are three hydrodynamic fields corresponding to the conservation of mass, momentum and energy. We provide physically motivated derivations of the hydrodynamic equations at the levels of Euler and then Navier-Stokes-Fourier.  Next we consider examples to test as to how successful the hydrodynamic description is in predicting the observed time evolution of nonequilibrium initial conditions such as domain walls and blasts. We find that in some cases there is good agreement of microscopic simulations with predictions from the Euler equations while, in several other cases, there is significant departure from the Euler predictions suggesting that the role of dissipation and noise is important in general.  
\end{abstract}
\tableofcontents

\section{Introduction}

 The question as to whether or not  an anharmonic chain with FPUT interactions reaches thermal equilibrium at long times has been studied extensively since the famous paper by FPUT in 1955~\cite{Fermi1955}. The issue is still debated and perhaps it is fair to say that the problem continues to be an open one. We point to several reviews that have appeared over the years~\cite{Ford1992,weissert1997,Berman2005,Gallavotti2008,Benettin2013}.

Another fascinating set of studies on the FPUT system is that on its transport properties. It was discovered by Lepri, Livi and Politi that the system exhibits anomalous transport~\cite{lepri1997} in the sense that  Fourier's law of heat conduction is not valid in this system. This conclusion follows  from studies in two physical frameworks~\cite{lepri1997,dhar2008,lepri2016}. In the first case one prepares a long FPUT chain in a Gibbs equilibrium state and sees how  perturbations spread or, equivalently, look at the form of space-time correlations of the conserved quantities~\cite{zhao2006,zaburdaev2011,chen2013,das2014}. For the FPUT chain it is found that perturbations typically spread super-diffusively. In the second framework one considers a nonequilibrium setup~\cite{lepri1997,mai2007,saito2010,das2014b} where a  FPUT chain with $N>>1$ particles is connected to heat baths (e.g. modeled through Langevin equations) at the two ends, with  the temperatures at the two ends differing by a small amount $\Delta T$.  It is then found that the heat current across the chain scales as $ N^{-\alpha} \Delta T$, with $0<\alpha<1$, which again indicates super-diffusive transport. 
 In  the two frameworks discussed above, the system is  either in equilibrium (by preparation), or, in a state of local equilibrium~\cite{mai2007}.   It is to be noted that anomalous transport is observed in a large class of momentum-conserving 1D systems, which include systems that have good thermalization properties. Hence, anomalous heat transport in the FPUT chain is unrelated to the original issue of its lack of thermalization. 
 
 A theoretical understanding of anomalous transport in anharmonic chains has been obtained via the formalism of fluctuating hydrodynamics discussed first by \cite{narayan2002,van2012} and developed systematically by \cite{mendl2013,spohn2014NFHT}. In particular, the theory makes detailed predictions for the form of dynamical correlations which have been tested with direct molecular dynamic simulations  of the FPUT chain~\cite{das2014} 
 and the agreement seems good in some parameter regimes. In spite of its successes, nonlinear fluctuating hydrodynamics is a phenomenological theory, and the question of rigorously establishing exponents and universality classes for the FPUT system is still open. There are surprising features that are  still not completely understood, for example the apparent long-time-length crossover to anomalous transport in some parameter regimes~\cite{zhao2012,chen2016,lepri2020}. 
 
With the above background, a natural question that arises is as to how good  the hydrodynamic description is, in describing far-from-equilibrium phenomena in the  FPUT chain.  The question of deriving hydrodynamics from microscopic dynamics is itself a fascinating and difficult problem (part of Hilbert's sixth problem) on which much progress has been made for the hard-sphere gas~\cite{gallagher2019}. However, taking as granted the validity  of the hydrodynamic equations, one can make predictions on the evolution of far-from-equilibrium initial states and ask how they compare with results from microscopic simulations. 
One of the interesting paradigmatic nonequilibrium initial conditions is the domain wall initial condition (equivalently the Riemann value problem, in the context of Euler equations). This was studied in ~\cite{Mendl_2016} for a 
number of 1D models where comparisons were made between the theoretical predictions of shocks, rarefactions and contact discontinuities that follow from Euler equations, with results for molecular dynamics. Recently, there have also been 
many interesting studies  on another far-from-equilibrium initial condition, namely the so-called blast problem~\cite{AKR2008,chakraborti2021,ganapa2021blast,Rajesh21a,Rajesh21b,chakraborti2022,singh2023,rajesh22,kumar2025}.  Here again, there have been detailed comparisons between the predictions of hydrodynamics with results from molecular dynamics. Rather surprisingly, it was observed~\cite{chakraborti2021} that the famous self-similar solutions of the Euler equations, obtained first by Taylor, von Neumann and Sedov (denoted as TvNS), are quite accurate even for the case of a one-dimensional gas of hard point particles. This is in spite of the fact that the 1D gas exhibits anomalous transport and has a divergent thermal conductivity. This motivates one to explore the validity of hydrodynamics in the FPUT system for the domain wall and blast initial conditions and this is the  main aim of this article. 

In this contribution we first present a review of known results on the hydrodynamic description of the FPUT chain. In general one can write hydrodynamics at three levels: (i) Euler hydrodynamics (ii) Navier-Stokes-Fourier (NSF) hydrodynamics and (iii) Noisy hydrodynamics. We discuss only the two levels and provide  heuristic derivations. Next we present numerical studies on the time evolution of two far-from-equilibrium initial conditions, namely a domain wall (or Riemann initial conditions) and a blast initial condition. We compare predictions from Euler hydrodynamics with observations from simulations. 

The plan of the next few sections is as follows. In Sec.~\eqref{sec:model} we define the FPUT Hamiltonian and state the hydrodynamic equations using both the fluid and lattice descriptions. In Sec.~\eqref{sec: derivations} we provide first the derivation of the Euler level hydrodynamic equations, again both in the fluid and lattice frameworks. Next we give two different derivations of the dissipative NSF equations, restricting now to the lattice formulation. As two applications of hydrodynamics in the far-from-equilibrium regime, we present in Sec.~\eqref{sec: numerics} the domain wall [Sec.~\eqref{subsec:dw}] and the blast initial  conditions [Sec.~\eqref{subsec:bp}], where we discuss first the theoretical formulations and present comparison between hydrodynamics and microscopic simulations. Finally we conclude with discussions in Sec.~\eqref{sec: summary}.

\section{The FPUT Hamiltonian and the hydrodynamic equations} 
\label{sec:model}
We consider  particles of unit masses on the real line with positions and momenta given by $q_X$ and $v_X$. We denote the interparticle separation between particles by $\hat r_X=\hat q_{X+1}-\hat q_{X}$. The  FPUT Hamiltonian is then given by:
\begin{align}
H&=\sum_{X \in \mathbb{Z}} \hat h_X,~~~ ~{\rm where}~~~\hat h_X=\frac{\hat v_X^2}{2}+V(\hat r_X),~~~~ 
V(r)=k_2\f{r^2}{2}+k_3\f{r^3}{3}+k_4\f{r^4}{4}.
\end{align}
The Hamiltonian equations of motion are given by:
\begin{align}
    \frac{d\hat q_X}{dt}=\p{H}/{\p \hat v_X}=\hat v_X,~~~~
    \frac{d\hat v_X} {dt}=-\p{H}/{\p \hat q_X}=V'(\hat r_X)-V'(\hat r_{X-1}),~~~X \in \mathbb{Z}.
\end{align}

It is expected that the only local conserved quantities for this system are the particle number (or mass), momentum and energy. Given these conservation laws, two different formulations of hydrodynamics are possible, based respectively  on the lattice and fluid views of the system. We now describe these formulations of hydrodynamics. In the context of anomalous transport in the FPUT chain, the lattice formulation turns out to be more useful. On the other hand, the fluid formulation is more common in the literature of fluid dynamics. Hence in this section we first present the fluid picture and then the lattice picture. For most of the later sections we will stick to the lattice formulation.

{\bf Fluid formulation of hydrodynamics:}
Viewing the FPUT system as a fluid, we expect that its hydrodynamics should  be described by the three coarse-grained conserved fields.  These are the mass density, $\rho(x,t)$, momentum density,  $p(x,t)$, and  energy density, $e(x,t)$. For a fluid, one defines a velocity field, $\vf(x,t)$, via the equations $p=\rho \vf$. The standard equations of fluid dynamics of a $1$D fluid, including Euler and dissipative terms are given by:   
\begin{subequations}
\label{eq:nsf}
 \begin{align}
&\f{\p \rho}{\p t}+\f{\p  (\rho \vf)}{\p x}=0, \label{eq:nsfa}\\ 
&\f{\p (\rho \vf)}{\p t}+\f{\p}{\p x}~(\rho \vf^2+ P) = {\f{\p }{\p x} \left(\eta \f{\p \vf}{\p x}\right)},  \label{eq:nsfb} \\
&\f{\p e}{\p t} +\f{\p }{\p x}~\left[ (e+P)\vf \right] ={\f{\p }{\p x} \left(\eta \vf \f{\p \vf}{\p x} + \kappa \f{\p T}{\p x} \right)}, 
\label{eq:nsfc}
\end{align}
\end{subequations}
where $P$ is the local pressure and $T$ the local temperature fields, that are related to the conserved fields via thermodynamics. The energy density has contributions from the centre of mass motion of the  fluid and from the internal energy, thus: 
\begin{align}
    e=\rho \left[\f{\vf^2}{2} + \varepsilon(\rho,T)\right],
\end{align}
where the internal energy per particle, $\varepsilon(\rho,T)$, is known from equilibrium thermodynamics. 
We can take either the temperature $T$ or the energy density as the independent hydrodynamic field. 
Similarly, the pressure is known in terms of the density and temperature from the equation of state 
\begin{align}
 P=P(\rho,T).   
\end{align}
The terms on the right hand side of Eqs.~\eqref{eq:nsf} are the dissipative NSF terms and involve the two transport coefficients, namely the bulk viscosity, $\eta$, and the thermal conductivity $\kappa$. In general, $\eta$ and $\kappa$  depend on  temperature and density.

{\bf Lattice formulation}: To obtain hydrodynamic equations in the lattice formulation from the fluid one, we  transform from the space fixed  coordinate, $x$, to  the particle coordinates $X$~\cite{spohn2014NFHT}. Noting the relation,
\begin{align}
    \int_{-\infty}^x dx' \rho(x') = X,
\end{align}
we get $\partial_x=\rho \p_X$. We also define the fluid derivative
\begin{align}
    \f{d}{dt}=\f{\p}{\p t}+\vf\f{\p}{\p x}.
\end{align}
Defining the stretch, $r_X=1/\rho(x)$,  the particle momentum (recalling that masses are unity), $\vf_X=\vf(x)$, and the energy per particle, $\e_X=e(x)/\rho(x)=v_X^2/2+\varepsilon_X$, as the lattice hydrodynamic fields, it is easily seen that the above equations take the form:
\begin{subequations}
\begin{align}
    &\f{d r}{d t}-\f{d \vf}{d X}=0, \\ 
&\f{d \vf}{d t}+\f{d P}{d X} = {\f{d }{d X} \left(\f{\eta}{r} \f{d \vf}{d X}\right)},  \\
&\f{d \e}{d t} +\f{d (\vf P)}{d X}~ ={\f{d }{d X} \left(\f{\eta \vf}{r} \f{d \vf}{d X} + \f{\kappa}{r} \f{d T}{d X} \right)}. 
\end{align}
\label{eq:nsf-lattice}
\end{subequations}
For comparison with the microscopic derivations in the next section, it is instructive to write the dissipative parts in terms of the gradients of the conserved fields. For this we use the relation $\e=\vf^2/2+\varepsilon(r,T)$  to arrive at the result:
\begin{align}
    \f{\eta \vf}{r} \f{d \vf}{d X} + \f{\kappa}{r} \f{d T}{d X} = \f{\kappa}{r} (\p_T \varepsilon)^{-1} \p_r \varepsilon \f{d r}{d X} - \f{\kappa}{r} (\p_T \varepsilon)^{-1}  \f{d \e}{d X} + \left[ \f{\kappa}{r} (\p_T \varepsilon)^{-1} +\f{\eta}{r} \right] \vf \f{d \vf}{d X}. 
    \label{eq:ediss}
\end{align}

\section{Phenomenological derivation of the hydrodynamic equations}
\label{sec: derivations}

One of the key ideas involved in developing hydrodynamic theory is  the idea of local equilibrium. This says that  a system with a large number of particles can be divided into spatial cells of some characteristic length, say $\ell$,  which is large compared to microscopic length scales but small compared to typical observational scales, such that within such cells, the state of the system is well described by specifying the few  conserved fields. Thus the system is in local equilibrium in such cells. We also make the assumption that the local fields change over time scales that are large compared to microscopic scales but sufficiently small compared to  observational times.  Let us denote by $\hu^\alpha_x(t)$, for $\alpha=1,2,3$, the empirical value of the conserved fields inside a cell centered at position $x$ and at time $t$.  

Our heuristic derivations of the FPUT hydrodynamics are based on the local equilibrium assumption and  following additional ideas:
\begin{itemize}
    \item Consider any particular realization of the empirical macroscopic fields $\hu^\alpha_x(t)$. There is an ensemble of microscopic configurations which would all correspond to the same value of these empirical fields. The assumption of local equilibrium means that at all times, this ensemble would be a local micro-canononical one which we denote as $P_{\rm LMC}(\Gamma)$, where  $\Gamma = \{ q_X,p_X\}$ denotes a point in phase space. Using the equivalence of ensembles, one can also work with a local-equilibrium Gibbs ensemble which we will denote as $P_{\rm LG}(\Gamma)$. 
\item We expect that for almost any initial microscopic configuration, drawn from $P_{\rm LMC}$, the empirical fields will exhibit the same time-evolution, with fluctuations which depend on the coarse-graining scale. This is the source of noise in the hydrodynamic equations. 
\item As we will demonstrate in the following sections, the Euler terms in the hydrodynamic equations follow simply by finding the averages of the microscopic currents in the local cells assumed to be in equilibrium. To obtain the dissipative terms we need to consider the time-evolution of the currents in local equilibrium states with small gradients in the conserved fields. Obtaining the noise terms requires more subtle computations for which we refer to ~\cite{saito2021}.    
\end{itemize}

\subsection{\bf Euler hydrodynamics}
\subsubsection{\bf Fluid formulation} 
We  define the empirical hydrodynamic fields in terms of the microscopic variables.
\begin{subequations}
\label{eq:mic-fl-fields}
\begin{align}
{\hat\rho(x,t)}&=\sum_{X}   \delta(x-\hat q_X(t)),~~~ \\
{\hat p(x,t)}&=\sum_{X}   \hat v_X ~\delta(x-\hat q_X(t)),  \\
{\hat e(x,t)}&=\sum_{X}   \hat h_X ~\delta(x-\hat q_X(t)). 
\end{align}
\end{subequations}
These satisfy the microscopic continuity equations:
\begin{align}
   & \p_t \hat \rho + \p_x \hat j^\rho =0,~~ \p_t \hat p + \p_x \hat j^p =0,~~\p_t \hat e + \p_x \hat j^e =0,~~
   \end{align}
where
\begin{subequations}
   \label{eq:mic-fl-fields}
\begin{align}
   & \hat j^\rho =  \sum_{X}   \hat v_X \delta(x-\hat q_X(t)) ,\\
   &\hat j^p =  \sum_{X}   \hat v^2_X \delta(x-\hat q_X(t)) + \hat F_X ~\theta(\hat q_X(t)-x), \\
   &\hat j^e =  \sum_{X}   \hat v_X\hat  h_X  \delta(x-\hat q_X(t)) + [ \hat v_{X+1} V'(\hat r_X) -\hat v_X V'(\hat r_{X-1})] ~  \theta(\hat q_X(t)-x).
\end{align}
\end{subequations}
Now consider a box of length $\ell$ which is  in local equilibrium  with mean density, $\rho$, mean momentum density, $p$ and mean  energy-density, $e$. We further define the mean velocity, $\vf$ and mean internal energy per particle, $\varepsilon$, through the relations:   
\begin{align}
    p=\rho \vf, ~~~~e=\rho (\vf^2/2 +\varepsilon). 
\end{align}
The local temperature, $T$, and pressure, $P$ in the box are obtained from the equilibrium  thermodynamic relations:
\begin{align}
\varepsilon=\varepsilon(\rho,T),~~~P=P(\rho,T).
\end{align}
The Euler currents are found by integrating the currents over the local equilibrium box of length $\ell$ and taking an average over the local microcanonical equilibrium ensemble, specified by the thermodynamic parameters $\rho,v,e$ and indicated below by the subscript MC. Thus we get:
\begin{subequations}
    \begin{align}
   &  j^\rho =  \f{1}{\ell} \sum_{\hat q_X \in (0,\ell)}   \left\la \hat v_X \right\ra_{\rm MC} = \rho \vf   ,\\
   & j^p =  \f{1}{\ell} \sum_{\hat q_X \in (0,\ell)}     \left\la [\hat v^2_X    + (V'(\hat r_X)-V'(\hat r_{X-1}))~ \hat q_X(t)] \right\ra_{\rm MC} = \rho \vf^2 + \f{1}{\ell} \sum_{\hat q_X \in (0,\ell)}     \left\la [(\hat v_X-\vf)^2    - V'(\hat r_X) \hat r_{X}~] \right\ra_{\rm MC}  , \nn \\
   &= \rho \vf^2 +P,\\
   & j^e =  \f{1}{\ell} \sum_{\hat q_X \in (0,\ell)}    \left\la \hat v_X \hat h_X   + [ \hat v_{X+1} V'(\hat r_X) -\hat v_X V'(\hat r_{X-1})] ~  \hat q_X(t) \right\ra_{\rm MC} \nn \\
   &=\f{1}{\ell} \sum_{\hat q_X \in (0,\ell)}    \left\la \f{\hat v_X^3}{2} + \hat v_X V(\hr_X)   -  \hat v_{X+1} V'(
   \hat r_X)  ~  \hat r_X(t) \right\ra_{\rm MC} \nn \\
  &=\f{1}{\ell} \sum_{\hat q_X \in (0,\ell)}     \f{\vf^3}{2} +\vf \left\la \f{(\hat v_X-\vf)^2}{2} +  V(\hr_X) \right\ra_{\rm MC} + \vf \left\la (\hat v_X-\vf)^2  -   V'(\hat r_X)  ~  \hat r_X \right\ra_{\rm MC}  \nn \\
  &=\vf(e+P),
\end{align}
\end{subequations}
where the identification $P=\f{1}{\ell} \sum_{\hat q_X \in (0,\ell)}     \left\la [(\hat v_X-\vf)^2    - V'(\hat r_X) \hat r_{X}~] \right\ra_{\rm MC}$ follows from the Virial theorem while the energy density is given by 
$e=\f{1}{\ell} \sum_{\hat q_X \in (0,\ell)}      v^2/2+ \left\la [(\hat v_X-\vf)^2/2  + V(\hat r_X)] \right\ra_{\rm MC}$.

\subsubsection{\bf Lattice formulation}
In the lattice formulation~\cite{spohn2014NFHT,lepri2016}, the conserved microscopic fields  are the stretch $\hr_X=\hat q_{X+1}-\hat q_X$, the velocity $\hv_X$ and the
energy per particle $\hh_X= \hv_X^2/2 + V(\hr_X)$. The corresponding hydrodynamic fields are $r_X = \la \hr_X\ra$, $\vf_X=\la \hat v_X \ra$, and $ \e_X=\la \hat h_X \ra =\vf_X^2/2+\varepsilon_X$, where $\varepsilon_X=\la (\hv_X-\vf_X)^2/2+ V(\hr_X) \ra$. Using the Hamilton equations of motion we get $\partial_t \hat r_X = \hat v_{X+1} - \hat v_X$, ~$\partial_t \hat p_X = V'(\hat r_X) - V'(\hat r_{X-1})$, and $\partial_t \hat h_X = \hat v_{X+1} V'(\hat r_X) - \hat v_X V'(\hat r_{X-1})$ and hence,
\begin{equation}
    \hat j^{\,r}_X = -\hat v_{X+1}, \qquad
    \hat j^{\,\vf}_X = -V'(\hat r_X), \qquad
    \hat j^{\,\e}_X = -\hat v_{X+1} V'(\hat r_X).
    \label{eq:micro_currents_lattice}
\end{equation}
Here, it is more convenient to take an average of observables over a Gibbs equilibrium state with specified inverse temperature, $\beta$, pressure, $P$, and velocity, $v$:
\begin{align}
    P_{\rm G}(\Gamma)=\f{1}{Z}e^{-\beta \sum_X \left(\hh_X -v \hv_X +P \hr_X \right) },~~~Z(P,\vf,\beta)=\int d \Gamma e^{-\beta\sum_X \left(\hh_X -v \hv_X +P \hr_X \right)}.
\end{align}
The single-site averages give $\langle \hat v_X \rangle_{\beta,P,\vf} = \vf$, and $\langle V'(\hat r_X) \rangle_{\beta,P,\vf} = -P$ and the equations of state are given by 
\begin{equation}
    r(\beta,P) = \langle \hat r_X \rangle_{\beta,P,\vf},
    \qquad
    \varepsilon(\beta,P)
        = \langle V(\hat r_X) \rangle_{\beta,P,\vf} + \frac{1}{2\beta}.
    \label{eq:eos_lattice}
\end{equation}
One can transform between the microcanonical (fixed $r,\vf,\e$) and Gibbs (fixed $P,\vf,\beta)$)  ensembles by using the above relations. The Euler current can be obtained by taking an average of the microscopic currents in Eq.~\eqref{eq:micro_currents_lattice}. We then get,
\begin{equation}
    j^{\,r} = -\vf, \qquad j^{\,\vf} = P, \qquad j^{\,\e} = \vf P,
    \label{eq:euler_currents_lattice}
\end{equation}
recovering the Euler part of Eqs.~\eqref{eq:nsf-lattice}.

\subsection{\bf Dissipative hydrodynamics}

In this section we only present the lattice formulation. We introduce the following notation for the microscopic fields:
\begin{align}
    \{\hu^{1}_X,~\hu^2_X,~\hu^3_X\} =\{ \hr_X,~\hv_X,~\hh_X~\},
\end{align}
and the corresponding currents $\{\hj^1_X,~\hj^2_X,~\hj^3_X\}$. 
The same variables without the hats denote the hydrodynamic fields. We will also use the notation for the conjugate hydrodynamic fields, that specify the local equilibrium state:  $\lambda^1_X=\beta_X,~\lambda^2_X=-\beta_X v_X,~\lambda^3_X=\beta_X P_X$.

\subsubsection{\bf Method I: evaluating current response to a small gradient in fields}
This approach for deriving the Green-Kubo type formulas was first discussed in ~\cite{visscher1974} and more recently in ~\cite{olla2019}
and, for quantum systems, in ~\cite{durnin2021}. To find the dissipative terms, we consider a state with small deviations from  equilibrium, one where the fields vary slowly in space  so that the phase-space distribution at some time ($t=0$) is of the local Gibbs form, 
\begin{equation}
P_{\rm LG}(\Gamma)
=\frac{1}{Z}\exp\!\left[
-\sum_{\alpha=1}^3 \sum_X \lambda^\alpha_X \hat{u}^\alpha_X\right],~~~Z=\int d \Gamma \exp\!\left[
-\sum_\alpha \sum_X \lambda^\alpha_X \hat{u}^\alpha_X\right].
\label{eqm dist}
\end{equation}
The slowly varying fields are taken to have  linear spatial dependence $\lambda_X^\alpha=\lambda^\alpha+X\epsilon^\alpha$, where the gradient, $\epsilon^\alpha=\p_X \lambda^\alpha|_{X=0}$, is assumed to be  small and a constant.  Then,  to leading order in the gradients,  $\epsilon^\alpha$, we get the local equilibrium distribution:
\begin{align}
P_{\rm LG}(\Gamma)
=\frac{1}{Z_0}\exp{\left[
-\sum_\alpha  \lambda^\alpha \sum_X \hat{u}^\alpha_X
\right]} \times \left[1  -\sum_\alpha  \epsilon^\alpha \sum_X X (\hat{u}^\alpha_X-u^\alpha) \right],~~~Z_0=\int d \Gamma \exp\!\left[
-\sum_\alpha \lambda^\alpha  \sum_X \hat{u}^\alpha_X\right].
\label{eqm dist le}
\end{align}
As seen in the previous section, the Euler current is simply given by the expectation value of the current operator, $\hj^\alpha$,  in the homogeneous Gibbs state $P_{\rm G}= \exp[-\lambda^\alpha \sum_X \hu^\alpha_X]/Z_0$. The Euler current is denoted by $j^\alpha$. We now  evaluate the excess current, $\Delta \hj^\alpha_0=\hj^\alpha_0-j^\alpha$,  at the site $X=0$ in the small gradient state given by Eq.~\eqref{eqm dist le}.
The expectation of $\Delta \hj^\alpha_0$ in this slowly changing local equilibrium state is given by:
\begin{align}
    \Delta j^\alpha_0&=-\sum_\beta \epsilon^\beta \sum_X X \left\langle \Delta \hat{j}^\alpha_0(t) (\hat{u}^\beta_X-u^\beta) \right\rangle_{\rm G},    \\
  {\rm hence,}~~\Delta j^\alpha_0  &=-\sum_\beta \epsilon^\beta \sum_X X \left\langle \Delta \hat{j}^\alpha_0(0)~ \left[\hat{u}^\beta_X(-t)-u^\beta\right] \right\rangle_{\rm G}, 
\end{align}
where $\la...\ra_{\rm G}$ denotes an average over the local homogeneous Gibbs state. From the continuity equation we get:
$\Delta \hu^\beta_X(-t)=\hat{u}^\beta_X(-t)-{u}^\beta = \Delta \hu^\beta_X(0) + \int_0^t ds [\Delta \hat{j}^\beta_{X+1}(-s) -\Delta \hat{j}^\beta_X(-s)]$ which leads to:
\begin{align}
 \Delta j^\alpha_0  &=-\sum_\beta \epsilon^\beta \sum_X  \left\{X\left\langle \Delta\hat{j}^\alpha_0(0) \int_0^t ds \left[\Delta \hat{j}^\beta_{X+1}(-s) -\Delta\hat{j}^\beta_X(-s)\right]\right\rangle_{\rm G}
 + X\left\langle
\Delta\hat{j}^\alpha_0(0)~  \Delta \hat{u}^\beta_X(0) \right\rangle_{\rm G} \right\} \\
&=\sum_\beta \epsilon^\beta \sum_X  \left\{ \int_0^t ds  \left\langle \Delta \hat{j}^\alpha_0(s) \Delta \hat{j}^\beta_{X}(0) \right\rangle_{\rm G}
 - X\left\langle
\Delta \hat{j}^\alpha_0(0)~  \Delta \hat{u}^\beta_X(0)  \right\rangle_{\rm G} \right\}
\end{align}
The second term vanishes, because of the product structure of the Gibbs state.  Let us denote the integrand in the first term by 
\begin{align}
\Gamma^{\alpha \beta}(s)=  \sum_X     \left\la \Delta \hat{j}^\alpha_{X}(s)\Delta \hat{j}^\beta_0(0) \right\ra_{\rm G}.  \label{eq:gammadef}
\end{align}
The long time limit, $\Gamma^{\alpha \beta}(s \to \infty)$, when non-vanishing, defines the Drude weight. Hence, taking the large $t$ limit,  we arrive at the following form for the current response:
\begin{align}
 \Delta j^\alpha_0  
&=\sum_\beta L^{\alpha \beta} \epsilon^\beta + t ~\Gamma^{\alpha \beta}(\infty) \epsilon^\beta,  \\
{\rm where}~~L^{\alpha \beta}&=
 \int_0^\infty dt ~\left[\Gamma^{\alpha \beta}(t)-\Gamma^{\alpha \beta}(\infty)\right], \label{eq:L}
\end{align}
defines the elements of the Onsager matrix, $L$. 
We now note that $\epsilon^\alpha = \p_X \lambda^\alpha=\p \lambda^\alpha/\p u^\beta \p_X u^\beta=-[C^{-1}]^{\alpha \beta} \p_Xu^\beta$, where $C$ is the static correlation matrix:
\begin{align}
    C^{\alpha \beta}= -\sum_X \left\la \Delta \hat{u}^\alpha_X(0)  \Delta\hat{u}^\beta_0(0) \right\ra_{\rm G}=-\f{\p u^\alpha}{\p \lambda^\beta}.\
\label{eq:C}
\end{align}
Using this we finally get the diffusion current
\begin{align}
  j^\alpha_{\rm Diff}=- D^{\alpha \beta} \p_X u^\beta,~~~
  {\rm where} ~~D=LC^{-1}, \label{eq:GKH}
\end{align}
is the diffusion matrix in the $P,v,\beta$ ensemble.

\subsubsection{\bf Method II: evolution of dynamical correlation functions}

 The second  approach we describe follows closely  that described in \cite{spohnbook} and also discussed recently in the context of generalized hydrodynamics~\cite{de2019}. One first establishes, from purely the microscopic dynamics, a relation between the spread of space-time correlations of the conserved fields with time integrals of temporal correlations of the conserved currents. Next, one considers  linearized hydrodynamic equations, with both Euler and dissipative terms, from which one can again compute the spread of correlations of the conserved fields and relate them to the drift and diffusion terms in the hydrodynamic equations. Comparing the results from the above two computations one arrives at expressions for the diffusion constant. We now give the details.

We define the space-time equilibrium correlation:
\begin{align}
  S^{\alpha \beta}_X(t)=\left\la \Delta \hat{u}^\alpha_X(t)  \Delta\hat{u}^\beta_0(0) \right\ra_{\rm G},
\end{align}
where $\Delta \hu^\beta_X(t)= \hu^\beta_X(t)- u^\beta$, and the average is taken in the Gibbs state specified by $P,u,\beta$. From the continuity equation one has $\Delta \hu^\alpha_X(t)=\Delta \hat{u}^\alpha_X(0)  -\int_0^t ds [\Delta \hat{j}^\alpha_{X+1}(s) -\Delta \hat{j}^\alpha_X(s)]$ which leads to:
\begin{align}
    &S^{\alpha \beta}_X(t)- S^{\alpha \beta}_X(0)=-\int_0^t ds \left\la 
    \left[\Delta \hat{j}^\alpha_{X+1}(s) -\Delta \hat{j}^\alpha_X(s)\right]
    \Delta\hat{u}^\beta_0(0) \right\ra_{\rm G} \nn \\
    &=-\int_0^t ds \left\la 
    \left[\Delta \hat{j}^\alpha_{X+1}(0) -\Delta \hat{j}^\alpha_X(0)\right]
    \Delta\hat{u}^\beta_0(-s) \right\ra_{\rm G} \nn \\
    &=-\int_0^t ds \left\la 
    [\Delta \hat{j}^\alpha_{X+1}(0) -\Delta \hat{j}^\alpha_X(0)] \int_0^s ds' [\Delta \hat{j}^\beta_{1}(-s') -\Delta \hat{j}^\beta_0(-s')]
    \right\ra_{\rm G} - t ~\left\la 
    [\Delta \hat{j}^\alpha_{X+1}(0) -\Delta \hat{j}^\alpha_X(0)]  \Delta \hu^\beta_0(0) \right\ra_{\rm G} \nn \\
    &=\int_0^t ds \int_0^s ds' \left[ \left\la \Delta\hat{j}^\alpha_{X+1}(s') \Delta\hat{j}^\beta_0(0) \right\ra -2 \left\la \Delta\hat{j}^\alpha_{X}(s') \Delta\hat{j}^\beta_0(0) \right\ra +\left\la \Delta\hat{j}^\alpha_{X-1}(s') \Delta \hat{j}^\beta_0(0) \right\ra   \right] 
\nn \\ 
&-t \left\la \left[ \Delta \hat{j}^\alpha_{X+1}(0)-\Delta \hat{j}^\alpha_X(0)\right] ~\Delta \hat{u}^\beta_0(0) \right\ra.
\end{align}
Taking care in defining the sums over $X$ in a symmetrized (about $X=0$) form, we then arrive at the following moments:
\begin{align}
\sum_X X S^{\alpha \beta}_X(t)&= B^{\alpha \beta} t = (A C)^{\alpha \beta} ~t\\
   \sum_X X^2 \left[S^{\alpha \beta}_X(t)-S^{\alpha \beta}_X(0)\right]&= 2 \int_0^t dt' \int_0^{t'} ds ~ \Gamma^{\alpha \beta}(s), \label{xsqexp1}
\end{align}
where the matrix $\Gamma$ is defined in Eq.~\eqref{eq:gammadef}, and $B$ is given by
\begin{align}
B^{\alpha \beta}= \sum_X  \la \hat{j}^\alpha_{X} (0)\hat{u}^\beta_0(0) \ra = (AC)^{\alpha \beta},  
\end{align}
where $C$ is defined in Eq.~\eqref{eq:C} and $A$ is the flux-Jacobian matrix
\begin{align}
    A^{\alpha \beta}= \f{\p j^\alpha}{\p u^\beta}.
\label{eq:A}
\end{align}
Let us again assume that the function $\Gamma^{\alpha \beta}(t)$ decays to a constant value at late times given by $\Gamma^{\alpha \beta}(\infty)$. Then at large times we can rewrite Eq.~\eqref{xsqexp1} in the form
\begin{align}
 \sum_X X^2 \left[S^{\alpha \beta}_X(t)-S^{\alpha \beta}_X(0)\right]&= 2 t L^{\alpha \beta} + t^2 \Gamma^{\alpha \beta}(\infty), \label{xsqexp2}
\end{align}
where the Onsager matrix $L$ is defined in Eq.~\eqref{eq:L}.

Next let us consider the predictions from  linearized  hydrodynamics according to which the correlation functions should evolve as:
\begin{align}
    \p_t S^{\alpha \beta}+  A^{\alpha \nu} \p_x S^{\nu \beta} =D^{\alpha \nu} \p_x^2 S^{\nu\beta}. 
\end{align}
This then leads to the moments evolving as:
\begin{align}
    \int dx ~x~ S^{\alpha \beta}  &= AC t \\
     \int dx ~x^2 ~ \left[ S^{\alpha \beta}(x,t)- S^{\alpha \beta}(x,0)\right] &= 2 (DC)^{\alpha \beta} t+ (A^2C)^{\alpha \beta}~ t^2 \label{eq:Shydro}
\end{align}
Finally comparing the expressions 
in Eqs.~\eqref{xsqexp2} and \eqref{eq:Shydro}, we recover the formula for the diffusion matrix $D=L C^{-1}$ derived earlier in Eq.~\eqref{eq:GKH}. In addition we obtain an equilibrium expression for the Drude weight:
\begin{align}
    \Gamma(\infty)=A^2C.
    \label{eq:drude}
\end{align}
After using the explicit expressions for $A$ and $C$, this has a simple form~\cite{spohnbook}.\\ 
\\
We now discuss the special case of the $v=0$ ensemble. In this case the Onsager matrix has a simple structure. Using the fact that the stretch current $\hj_X^r=\hv_X$ is itself a conserved variable, it follows immediately that the matrix elements $L^{1,\alpha}=L^{\alpha,1}$, for $\alpha=1-3$,  vanish identically. Secondly, noting that the momentum current, $\Delta \hj^v$, and energy current, $\Delta \hj^\e$, have opposite parities under time reversal (in the $v=0$ case), it follows that $L^{23}=L^{32}=0$. Thus there are only two non-vanishing coefficients $L^{22}$ and $L^{33}$. The correlation matrix in the $P,\beta$ ensemble is readily computed and given by:
\begin{align}
    C=\begin{pmatrix}
\la (\Delta \hr)^2\ra_{P,\beta} & 0 & \la \Delta \hr \Delta V(\hr) \ra_{P,\beta}\\
0 & \beta^{-1} & 0\\
 \la \Delta \hr \Delta V(\hr) \ra_{P,\beta} & 0 & \la(\Delta \hat h)^2  \ra_{P,\beta} 
\end{pmatrix}.
\end{align}
Using the result $D=LC^{-1}$ then gives us the following non-zero elements of the diffusion matrix:
\begin{align}
    D^{22}=\beta L^{22},~~
    D^{31}=\f{\la \Delta \hr \Delta  V(\hr) \ra_{P,\beta}}{\la \Delta \hr \Delta  V(\hr) \ra_{P,\beta}^2-\la (\Delta \hr)^2 \ra_{P,\beta} \la (\Delta \hat h)^2 \ra_{P,\beta}} L^{33},~~
    D^{33}=-\f{\la (\Delta \hr)^2 \ra_{P,\beta}}{\la \Delta \hr \Delta  V(\hr) \ra_{P,\beta}} D^{31} 
\end{align}
\begin{figure}[t]
    \centering
    \includegraphics[width=0.492\linewidth]{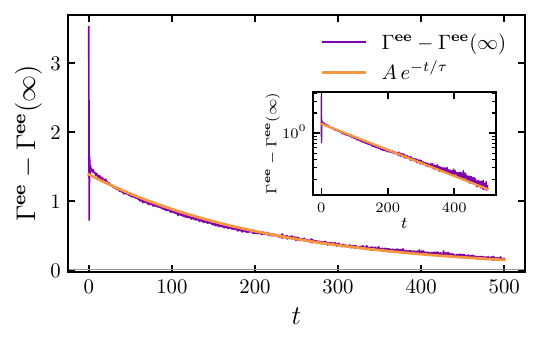}
    \includegraphics[width=0.48\linewidth]{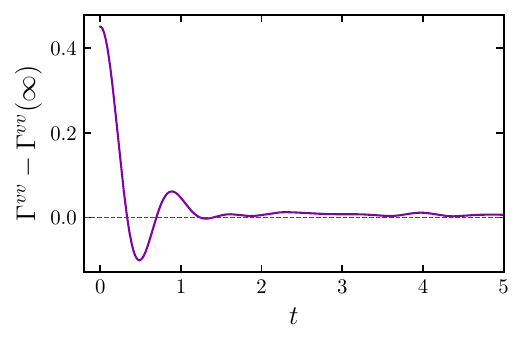}
    \caption{Current-correlators with the theoretical Drude weight values subtracted for $N=1000$, $\beta=1$, $P=2$, $v=0$. Left: $\Gamma^{\e\e}(t)-\Gamma^{\e\e}(\infty)$ (purple) with exponential fit (orange); inset shows the same on a semilog scale. Right: $\Gamma^{vv}(t)-\Gamma^{vv}(\infty)$.}
    \label{fig:correlators}
\end{figure}
We note that the ratio of the coefficients of the first two gradient terms on the right side of Eq.~\eqref{eq:ediss} is given by $-\p \varepsilon/\p r=-\p \la V(\hr)\ra_{r,T}/\p r=-\f{\p \la V(\hr)\ra_{P,T}/\p P}{\p \la \hr\ra_{P,T}/\p P}$ and agrees with the above microscopic computation. As expected, in the $v=0$ ensemble, the $vdv/dX$ term in Eq.~\eqref{eq:ediss} does not appear. 

For illustration, we now present some results from simulations for the current-current correlators. We numerically verify that  indeed all matrix elements of $\Gamma^{\alpha\beta}(t) -\Gamma^{\alpha\beta}(\infty)$ are zero except $\Gamma^{vv}(t)-\Gamma^{vv}(\infty)$ and $\Gamma^{\e \e}(t)-\Gamma^{\e \e}(\infty)$. In Fig.~\ref{fig:correlators} we show plots of these functions obtained from numerical simulations of the FPUT chain with parameters $(k_2,k_3,k_4)=(1,1,1)$ and $\beta=1$, $P=2$ where we average over $10^5$ and $2\times 10^5$ realizations respectively for the energy-energy and velocity-velocity current correlators. The Drude weights $\Gamma(\infty)$ were obtained using Eq.~\eqref{eq:drude} and the simplified expression given in~\cite{spohnbook}. From the numerics, we see that the energy current has an exponential decay though seeing the saturation to the expected Drude weight will take much longer times. It is also expected that at  longer time scales, one should see a cross-over to a power-law decay regime (see e.g. \cite{zhao2012,chen2016,lepri2020}). On the other hand, for the momentum  correlations we see a rapid oscillatory decay. 
\section{Examples of evolution of nonequilibrium initial states}
\label{sec: numerics}
\subsection{\textbf{Domain Wall initial conditions}}
\label{subsec:dw}
We now consider the evolution of a domain wall, or Riemann, initial condition for the FPUT chain in the lattice coordinate \(X\). At the hydrodynamic level, the initial data are taken to be piecewise constant:
\begin{equation}
(r,\vf,\varepsilon)(X,0)=
\begin{cases}
(r_L,\vf_L,\varepsilon_L), & X<0,\\[0.3em]
(r_R,\vf_R,\varepsilon_R), & X>0.
\end{cases}
\label{eq:DW-riemann-data}
\end{equation}
Here \(r(X,t)\) is the stretch field, \(\vf(X,t)\) is the velocity field, and \(\varepsilon(X,t)\) is the internal energy per particle. The left and right states are assumed to be equilibrium states, and we denote their corresponding thermodynamic pressures by $P_L \equiv P(r_L,\varepsilon_L)$ and $P_R \equiv P(r_R,\varepsilon_R)$. Likewise, the corresponding inverse temperatures are denoted by $\beta_L \equiv \beta(r_L,\varepsilon_L)$, and $\beta_R \equiv \beta(r_R,\varepsilon_R)$.

To compare the hydrodynamic predictions with microscopic simulations, we prepare the chain from a local-equilibrium measure which is different on the two sides of the origin. For given \((\beta,P,\vf)\), the equilibrium measure for one half-chain is taken to be
\begin{equation}
d\mu_{\beta,P,\vf}
\propto
\prod_X
\exp\!\left[
-\beta
\left(
\frac{(\hat v_X-\vf)^2}{2}+V(\hat r_X)+P\hat r_X
\right)
\right]
\, d\hat r_X \, d\hat v_X .
\label{eq:DW-local-eq-measure}
\end{equation}
Thus, on the left half \(X<0\) we sample from \(d\mu_{\beta_L,P_L,\vf_L}\), while on the right half \(X>0\) we sample from \(d\mu_{\beta_R,P_R,\vf_R}\). In practice, this means that the microscopic stretch and velocity variables are sampled independently in each half from the corresponding equilibrium distribution, after which the positions are constructed recursively from the stretches and the full chain is evolved under the Hamiltonian dynamics using RK4 with time step $10^{-3}$. In all the simulations that follow we check that the total energy and momentum drifts are negligible.

The thermodynamic fields appearing in Eq.~\eqref{eq:DW-riemann-data} are the equilibrium averages corresponding to the parameters \((\beta_{L,R},P_{L,R},\vf_{L,R})\).
In our simulations we take $\vf_L=\vf_R=0$ and for this section use the FPUT parameters $k_2=k_3=k_4=1$. 

The problem is then the following: starting from the microscopic initialization described above, we evolve the Hamiltonian dynamics and compare the resulting ensemble averaged profiles with the hydrodynamic solution of the corresponding Riemann problem. In secs.~\eqref{subsec:domain_wall_euler}, \eqref{sec:construct}, we describe the method of solution of the Riemann problem, closely following ~\cite{Mendl_2016,leveque1992numerical}. Then in Secs.~\eqref{sec:uneq_pressure}, \eqref{sec:eq-pressure}, we present the comparison with microscopic simulations.

\subsubsection{\bf Euler equations and elementary wave solutions}
\label{subsec:domain_wall_euler}

At the Euler scale, the hydrodynamic fields satisfy
\begin{subequations}
\begin{align}
&\partial_t r - \partial_X \vf = 0,\\
&\partial_t \vf + \partial_X P(r,\varepsilon) = 0,\\
&\partial_t \e + \partial_X (\vf  P(r,\varepsilon)\,) = 0.
\end{align}
\label{eq: euler_dw}
\end{subequations}

The 3rd equation in terms of the internal energy per particle $\varepsilon$ becomes
\begin{equation}
    \partial_t \varepsilon + P(r,\varepsilon)\,\partial_X \vf = 0.
\end{equation}

For the domain-wall initial condition Eq.~\eqref{eq:DW-riemann-data}, the Euler equations contain no intrinsic macroscopic length scale. The initial data are piecewise constant, with a single discontinuity at \(X=0\), and the Euler theory itself contains only first-order derivatives in space and time. It is therefore natural to seek a self-similar solution of the form $(r,\vf,\varepsilon)(X,t)=(r,\vf,\varepsilon)(\xi)$, where $\xi={X}/{t}$. Another way to see this is that the Riemann initial data are invariant under the rescaling $(X,t)\mapsto (\lambda X,\lambda t)$, and the Euler equations preserve this scaling. One therefore expects the solution to inherit the same symmetry, which forces it to depend on \(X\) and \(t\) only through the combination \(X/t\). Substituting this scaling form into Eqs.~\eqref{eq: euler_dw} gives a system of ordinary differential equations in \(\xi\), but a more useful way to proceed is to analyze the characteristic structure of the Euler system and then construct the solution from the corresponding elementary waves.

The structure of solutions to the Riemann problem is determined by the characteristic structure of the Euler system (see \cite{Mendl_2016,leveque1992numerical}). Writing the equations in quasilinear form,
\begin{equation}
\p_t\cf + A(\cf)\,\p_X\cf = 0, \qquad \cf=
\begin{pmatrix}
r\\
\vf\\
\e
\end{pmatrix}
\label{quasi-linear}
\end{equation}
one studies the spectrum of the flux Jacobian matrix \(A(\cf)\), defined in Eq.~\eqref{eq:A}. For each state \(\cf\), the eigenvalues \(\lambda_k(\cf)\) of \(A(\cf)\) are the characteristic speeds, and the corresponding eigenvectors \(\Psi_k(\cf)\) determine the directions of the \(k\)-th wave family in state space. These characteristic fields form the basic building blocks for the construction of the three types of elementary waves, namely, shocks, rarefaction waves, and contact discontinuities.
\\
For FPUT, it is easy to see that the flux Jacobian matrix is
\begin{equation}
A(\cf)=
\begin{pmatrix}
0 & -1 & 0\\
\partial_r P(r,\varepsilon) & -\vf\partial_{\varepsilon}P(r, \varepsilon) & \partial_\varepsilon P(r,\varepsilon)\\
\vf\partial_{r}P(r, \varepsilon) & P(r,\varepsilon)-\vf^2\partial_{\varepsilon}P(r, \varepsilon) & \vf\partial_{\varepsilon}P(r, \varepsilon)
\end{pmatrix}.
\label{eq:DW-jacobian}
\end{equation}
The eigenvalues are
\begin{equation}
\lambda_{-1}(r,\varepsilon)=-c(r,\varepsilon),
\qquad
\lambda_0(r,\varepsilon)=0,
\qquad
\lambda_1(r,\varepsilon)=c(r,\varepsilon),
\label{eq:DW-eigenvalues}
\end{equation}
where the sound speed is given by
\begin{equation}
c(r,\varepsilon)^2
=
P(r,\varepsilon)\,\partial_\varepsilon P(r,\varepsilon)
-
\partial_r P(r,\varepsilon).
\label{eq:DW-sound-speed}
\end{equation}
Thus the Euler system has three wave families: a left-moving acoustic family, a stationary middle family, and a right-moving acoustic family.

A convenient choice of right eigenvectors is
\begin{equation}
\Psi _{-1}=
\begin{pmatrix}
1\\
c\\
-P+\vf c
\end{pmatrix},
\qquad
\Psi_0=
\begin{pmatrix}
\partial_\varepsilon P\\
0\\
-\partial_r P
\end{pmatrix},
\qquad
\Psi_1=
\begin{pmatrix}
-1\\
c\\
P+\vf c
\end{pmatrix},
\label{eq:DW-eigenvectors}
\end{equation}
where \(P=P(r,\varepsilon)\) and \(c=c(r,\varepsilon)\).\\
\\
\noindent \textbf{Rarefaction waves:}
A \(k\)-rarefaction wave is a smooth self-similar solution for which the state varies continuously inside a region in the $X-t$ plane. Substituting \(\cf=
\cf(\xi)\) into Eq.~\eqref{quasi-linear} gives
\begin{equation}
\bigl(A(\cf)-\xi I\bigr)\,\f{d\cf(\xi)}{d \xi}=0.
\label{eq:DW-rarefaction-eq}
\end{equation}
Hence, in a \(k\)-rarefaction, the derivative \(\cf'(\xi)\) must be parallel to the \(k\)-th eigenvector, and one may parametrize the wave curve by
\begin{equation}
\frac{d\cf}{d\tau} = \Psi_k(\cf(\tau)),
\qquad
\xi = \lambda_k(\cf(\tau)).
\label{eq:DW-rarefaction-curve}
\end{equation}
Starting from a reference state \(\cf_a\), integration of Eq. \eqref{eq:DW-rarefaction-curve} generates the \(k\)-th rarefaction curve issuing from \(\cf_a\). For the FPUT Euler system, the first and third families are the acoustic families. It is convenient to label them by $\lambda_\sigma(r,\varepsilon)=\sigma\, c(r,\varepsilon)$, with $\sigma=\pm 1$. A corresponding right eigenvector may be chosen as
\begin{equation}
\Psi_\sigma(r,\vf,\varepsilon)=
\begin{pmatrix}
-\sigma\\
c(r,\varepsilon)\\
\sigma P(r,\varepsilon)+\vf c
\end{pmatrix}.
\end{equation}
Hence the integral curves of the acoustic rarefaction families satisfy
\begin{equation}
\frac{d}{d\tau}
\begin{pmatrix}
r\\
\vf\\
\e
\end{pmatrix}
=
\begin{pmatrix}
-\sigma\\
c\\
\sigma P+\vf c
\end{pmatrix},
\qquad \sigma=\pm 1.
\label{eq:fput-rarefaction-tau}
\end{equation}
Equivalently, eliminating the parameter \(\tau\), one obtains
\begin{equation}
\frac{d\vf}{dr}=-\sigma\, c(r,\varepsilon),
\label{eq:fput-rarefaction-u-r}
\end{equation}
\begin{equation}
\frac{d\varepsilon}{dr}=-P(r,\varepsilon).
\label{eq:fput-rarefaction-e-r}
\end{equation}
These two equations determine the rarefaction curves in the \((r,\vf,\varepsilon)\)-state space once the equation of state \(P(r,\varepsilon)\) is known. Inside the rarefaction fan, the similarity variable is related to the local state by $\xi=\lambda_\sigma(\cf)=\sigma\, c(r,\varepsilon).$ Thus a \(\sigma\)-rarefaction connecting a left state \(\cf_L\) to a right state \(\cf_R\) occupies the wedge $\lambda_\sigma(\cf_L)\le \frac{X}{t}\le \lambda_\sigma(\cf_R)$, within which the state varies smoothly along the integral curve of Eq. \eqref{eq:fput-rarefaction-tau}. Outside this region, the solution remains constant.\\
\\
\noindent \textbf{Shock waves:} A shock is a discontinuous weak solution of the conservative system Eq.~\eqref{eq: euler_dw}. Suppose a discontinuity moves with speed \(s\) and connects a left state \(\cf_a=(r_a,\vf_a,\e_a)\) to a right state \(\cf_b=(r_b,\vf_b,\e_b)\). Then the Rankine-Hugoniot conditions \cite{landau1987fluid} read
\begin{subequations}
\begin{align}
&s(r_b-r_a)=-(\vf_b-\vf_a),\\
&s(\vf_b-\vf_a)=P_b-P_a,\\
&s(\e_b-\e_a)=\vf_bP_b-\vf_aP_a,
\end{align}
\label{eq: RH_dw}
\end{subequations}
where \(P_a=P(r_a,\varepsilon_a)\) and \(P_b=P(r_b,\varepsilon_b)\) and $s$ is the shock speed. Equivalently, one may write $s[ r ] = -[ \vf ]$, $s[ \vf ] = [ P ]$, $s[ \e ] = [ \vf P ]$, where \([f]=f_b-f_a\) denotes the jump across the shock. The above equations lead to the following  simpler equations,
\begin{subequations}
\begin{align}
[\varepsilon] &= -\frac{1}{2}[r]\,(P_a+P_b), \\
-[\vf]^2 &= [r]\,[P],
\end{align}
\label{eq: RH_dwp}
\end{subequations}
which can be used to construct the shock curves in the $(r,v,\varepsilon)$ state space. \\
\\
\textbf{Contact Discontinuity:} Because the second characteristic speed is identically zero, the middle wave is a stationary contact discontinuity. Setting \(s=0\) in Eqs. \eqref{eq: RH_dw} gives $[\vf]=0$, and $[P]=0$. Thus the middle wave connects two states with the same velocity and the same pressure, while the stretch and internal energy may jump. This fact is central in the construction of the full Riemann solution.\\
\\
Having described the three elementary waves --- rarefactions, shocks, and contact discontinuities --- we now state the admissibility conditions that select the physically relevant branches. For a \(k\)-rarefaction, the solution must spread out into a fan, which requires that the \(k\)-th characteristic speed increase across the wave:
\begin{equation}
\lambda_k(\cf_L) < \lambda_k(\cf_R).
\label{rf-condition}
\end{equation}
For a \(k\)-shock, admissibility is imposed by the Lax entropy condition, namely
\begin{equation}
\lambda_k(\cf_L) > s > \lambda_k(\cf_R).
\label{lax-condition}
\end{equation}
Thus, in an admissible shock, characteristics of the \(k\)-th family impinge on the discontinuity from both sides, whereas in an admissible rarefaction they spread apart.

\subsubsection{\bf Construction of the full Riemann solution}
\label{sec:construct}
We can now describe the standard construction of the Euler solution of the domain-wall problem. Starting from the left state \(\cf_L=(r_L,\vf_L, \e_L)\), we follow the $-1$-family wave curve, allowing either a $-1$-rarefaction or a $-1$-shock. Starting from the right state \(\cf_R=(r_R,\vf_R, \e_R)\), we similarly follow the $1$-family wave curve, allowing either a $1$-rarefaction or a $1$-shock. The solution is obtained by finding two intermediate states, $\cf_{L*}=(r_{L*},\vf_*,\e_{L*}),$ and $\cf_{R*}=(r_{R*},\vf_*, \e_{R*}),$ such that
\begin{equation}
\cf_{L*}\in \mathcal{W}_{-1}(\cf_L),
\qquad
\cf_{R*}\in \mathcal{W}_{1}(\cf_R),
\qquad
\vf_{L*}=\vf_{R*}=\vf_*,
\qquad
P(\cf_{L*})=P(\cf_{R*})=P_*.
\label{eq:DW-star-matching}
\end{equation}
Here \(\mathcal{W}_{-1}(\cf_L)\) denotes the $-1$-family wave curve issuing from \(\cf_L\), while \(\mathcal{W}_1(\cf_R)\) denotes the $1$-family wave curve issuing from \(\cf_R\). The two intermediate states are then connected by the stationary contact discontinuity of the 0-family. Therefore, the full Euler solution consists of three elementary waves:
\begin{equation}
\cf_L
\;\xrightarrow{\;-1\text{-wave}\;}\;
\cf_{L*}
\;\xrightarrow{\;0\text{-contact}\;}\;
\cf_{R*}
\;\xrightarrow{\;1\text{-wave}\;}\;
\cf_R.
\label{eq:DW-wave-pattern}
\end{equation}
Depending on the initial left and right states, the first and third waves may each be either shocks or rarefactions. Since this wave pattern is not known in advance, one first constructs candidate solutions by allowing the outer waves to be either shocks or rarefactions and solving for the corresponding intermediate states. The physically relevant solution is then selected by imposing the appropriate admissibility conditions: rarefactions must have increasing characteristic speed across the wave Eq.~\eqref{rf-condition}, while shock branches must satisfy the Lax entropy condition Eq.~\eqref{lax-condition}.

\subsubsection{\bf Unequal-pressure domain wall}
\label{sec:uneq_pressure}
We now illustrate the above construction for a domain-wall initial condition with unequal pressures on the two sides. We take $P_L=P(r_L,\varepsilon_L)$, $P_R=P(r_R,\varepsilon_R)$, with $P_L \neq P_R $.

At the microscopic level, the left and right halves are sampled independently from equilibrium measures with parameters \((\beta_L,P_L,0)\) and \((\beta_R,P_R,0)\), respectively, and the resulting initial condition is evolved under the Hamiltonian dynamics. While solving the Euler equations for FPUT, the equation of state cannot be determined exactly and hence we calculate it numerically. We do this by solving the system $\{r = \langle \hat r \rangle_{P,\beta}, \quad \varepsilon = \frac{1}{2\beta} + \langle V(\hat r) \rangle_{P,\beta}\}$ 
numerically via integration and root-finding to extract $(P, \beta)$ from the macroscopic observables. For the particular initial condition considered here, the admissibility conditions select the wave pattern
\begin{equation}
\cf_L
\;\xrightarrow{\;\text{-1-shock}\;}
\cf_{L*}
\;\xrightarrow{\;\text{contact discontinuity}\;}
\cf_{R*}
\;\xrightarrow{\;\text{1-rarefaction}\;}
\cf_R .
\label{eq:uneq-wave-pattern}
\end{equation}

\begin{figure}[t]
%\centering
\includegraphics[width=0.24\textwidth]{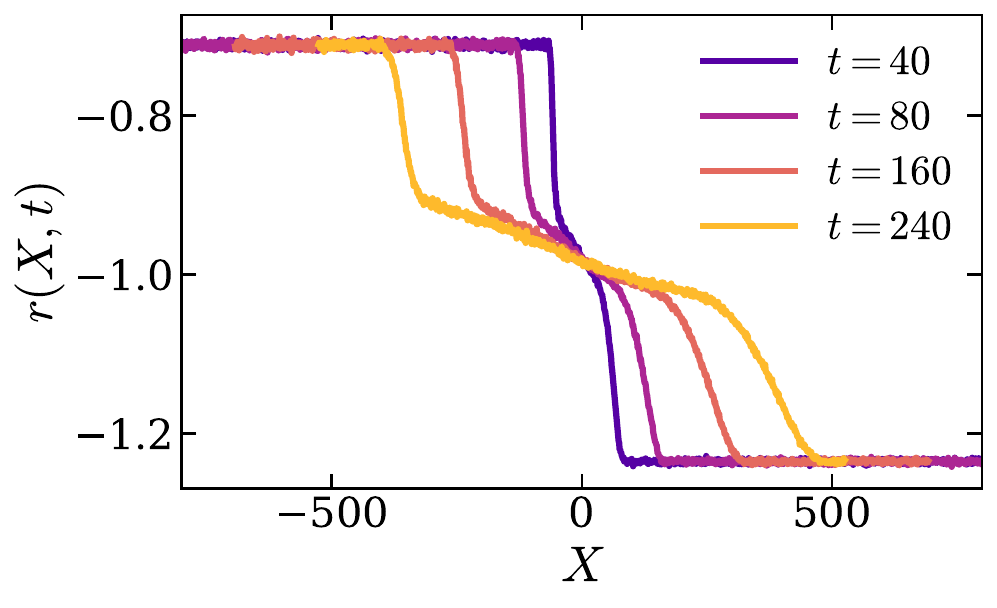}
\includegraphics[width=0.24\textwidth]{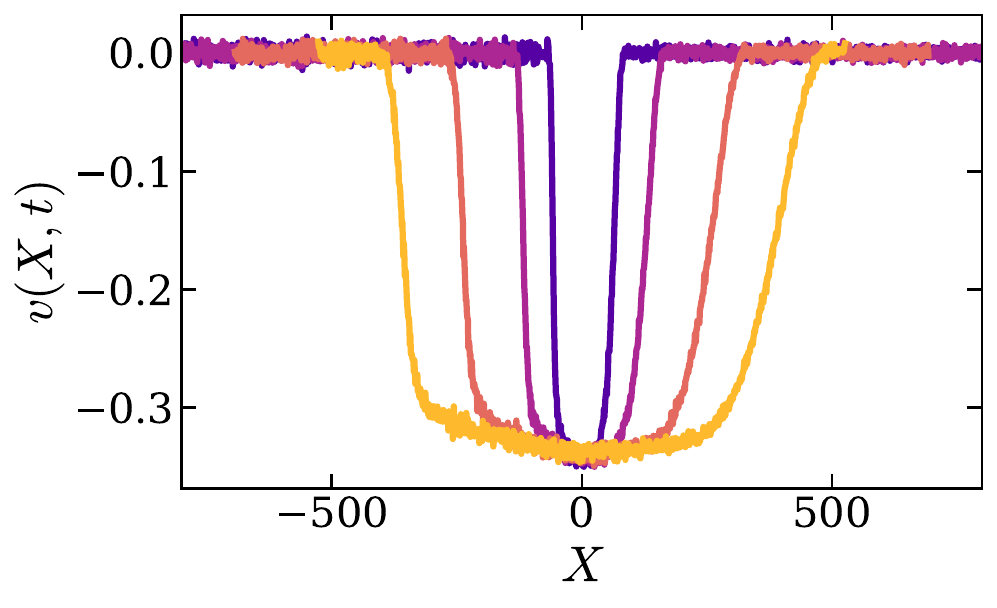}
\includegraphics[width=0.24\textwidth]{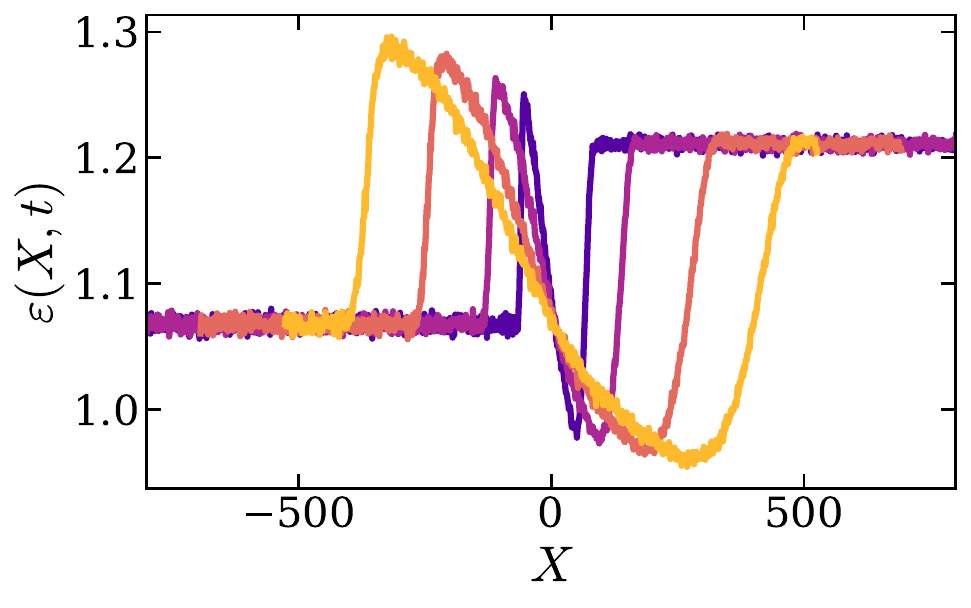}
\includegraphics[width=0.24\textwidth]{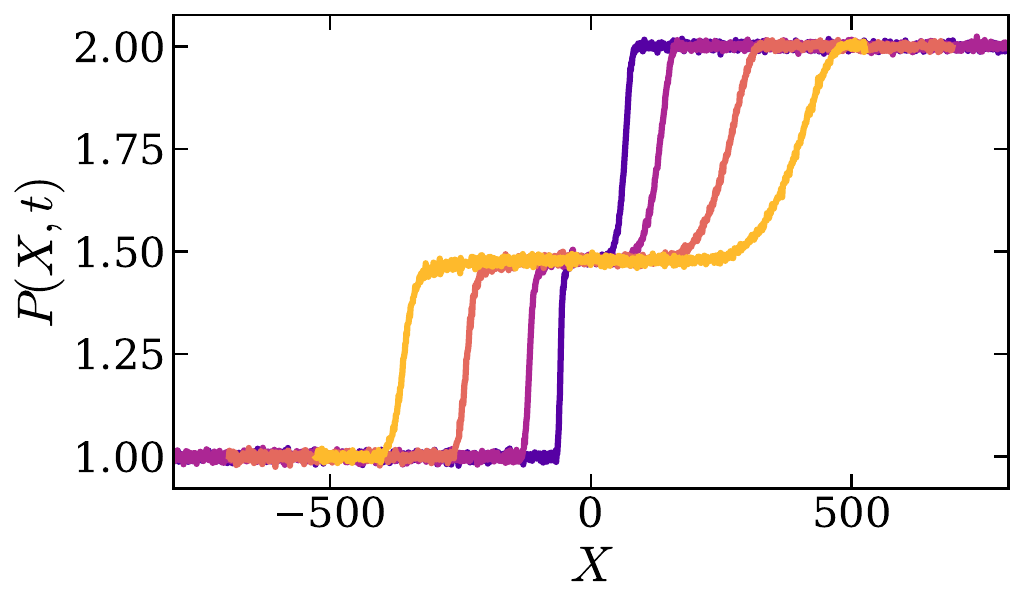}
%\caption{
%Profiles.
%\label{fig:uneq_pressure_profiles}
%\end{figure}
%\begin{figure}[t]
%\centering
\includegraphics[width=0.24\textwidth]{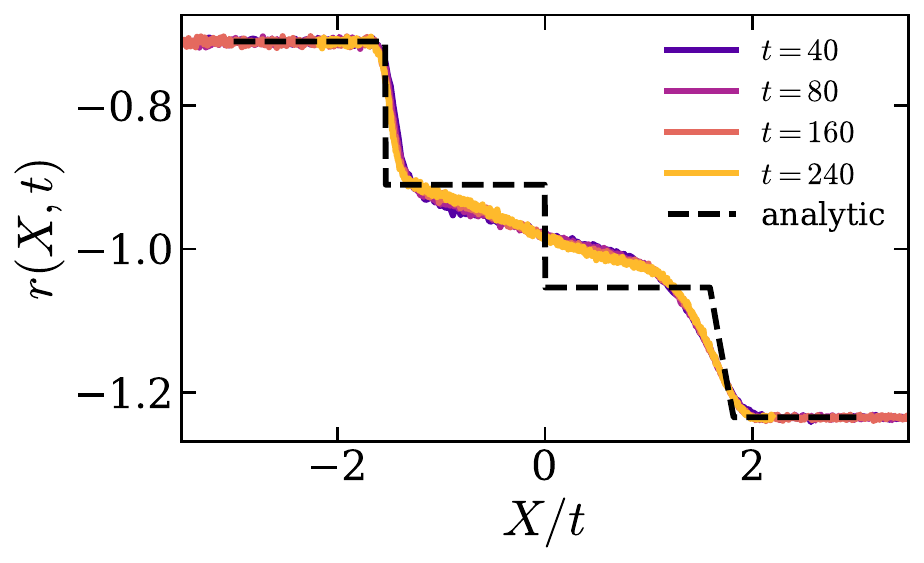}
\includegraphics[width=0.24\textwidth]{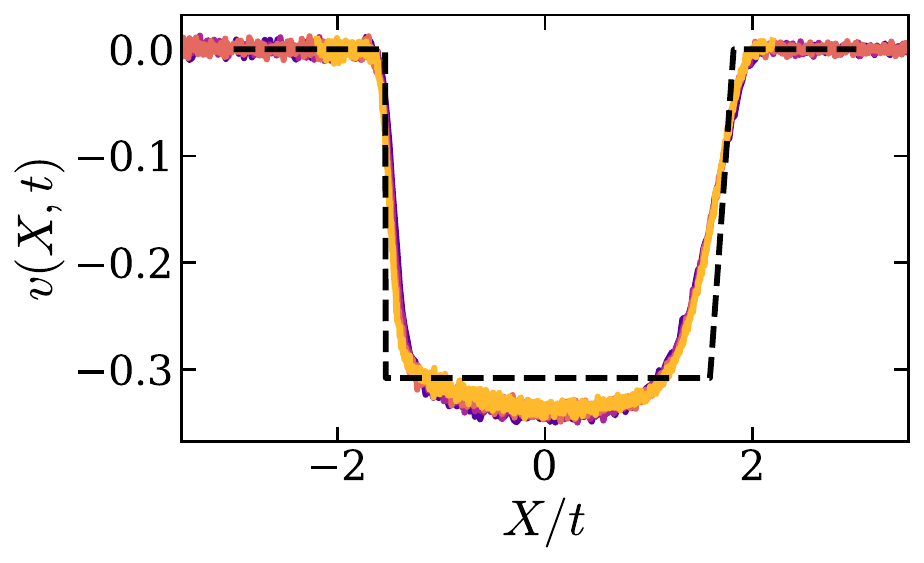}
\includegraphics[width=0.24\textwidth]{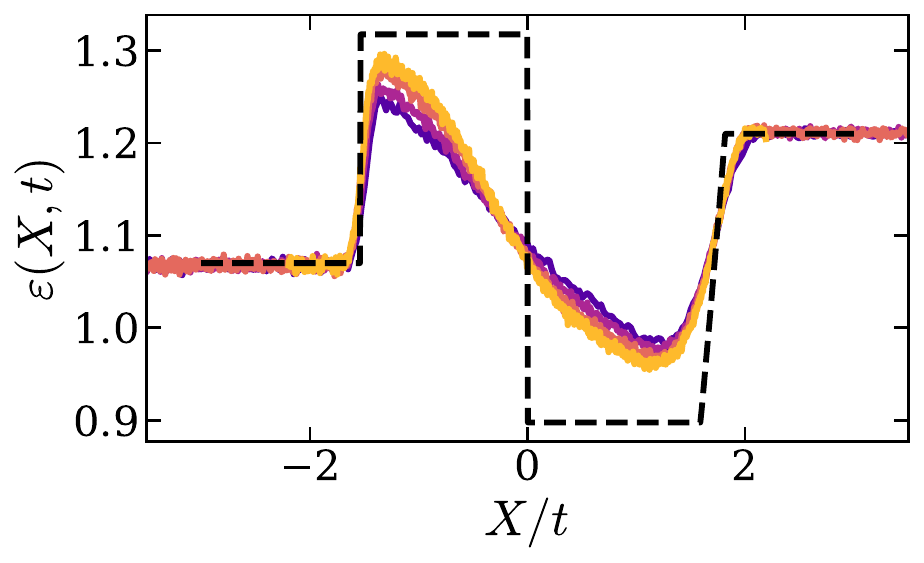}
\includegraphics[width=0.24\textwidth]{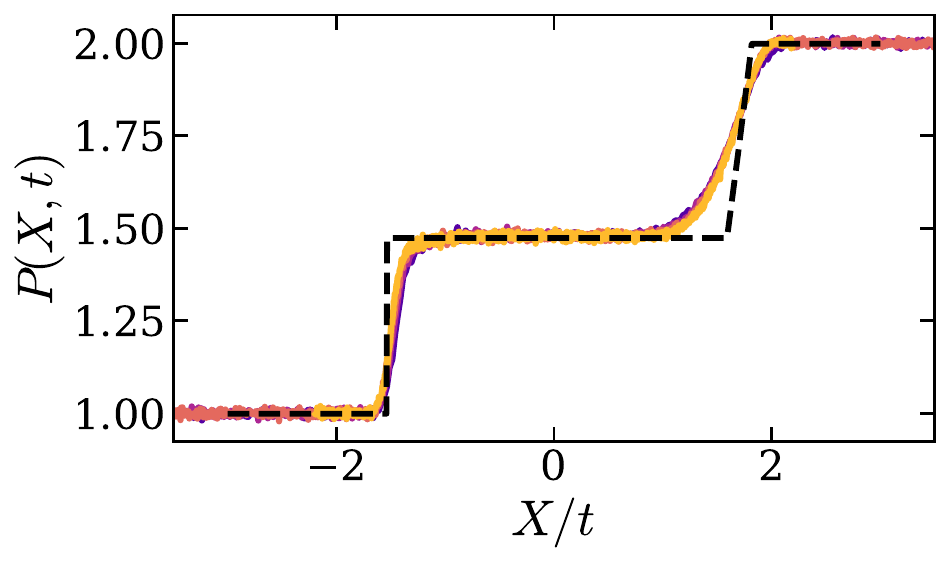}
\includegraphics[width=0.24\textwidth]{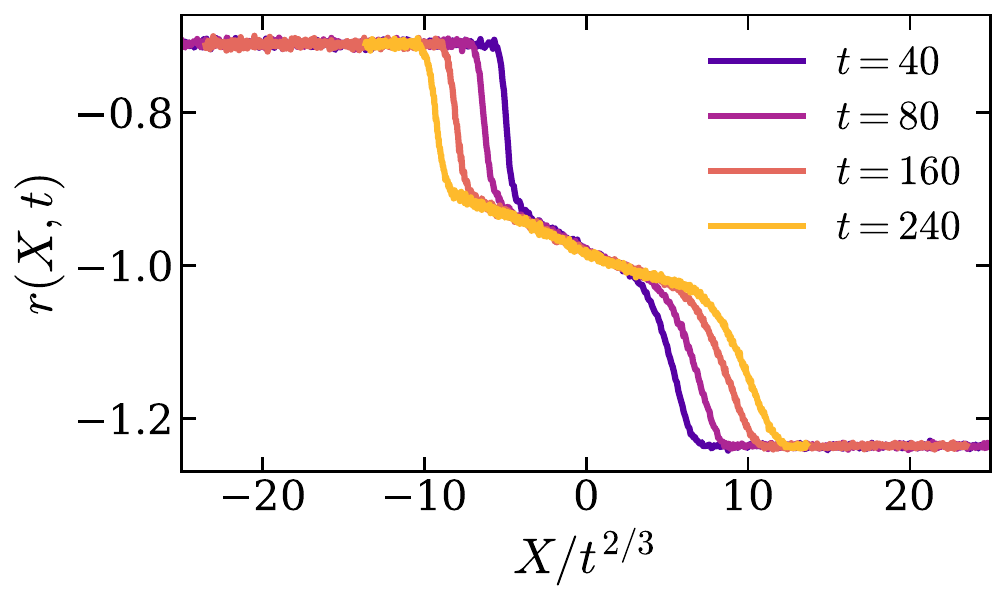}
\includegraphics[width=0.24\textwidth]{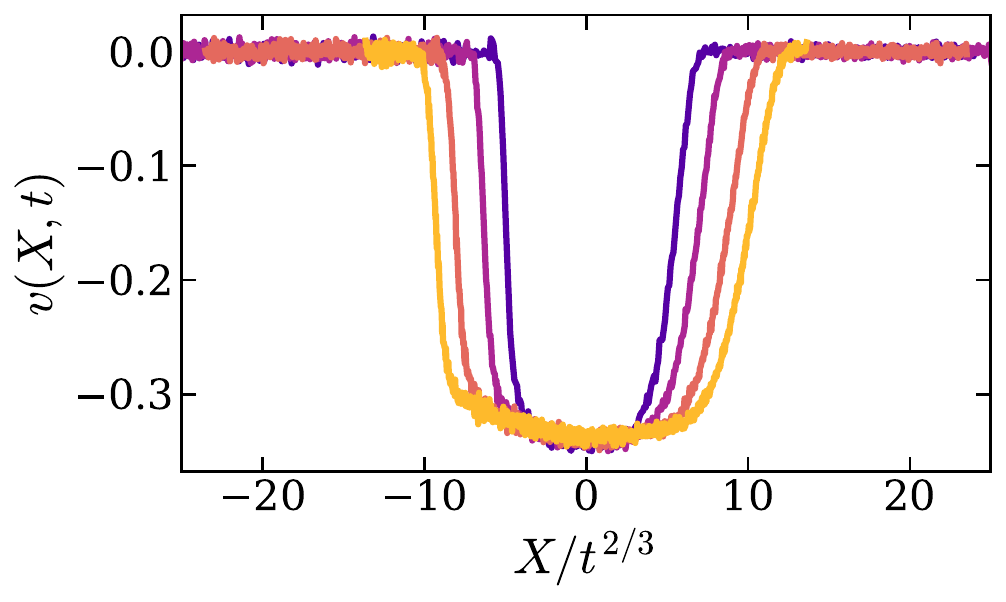}
\includegraphics[width=0.24\textwidth]{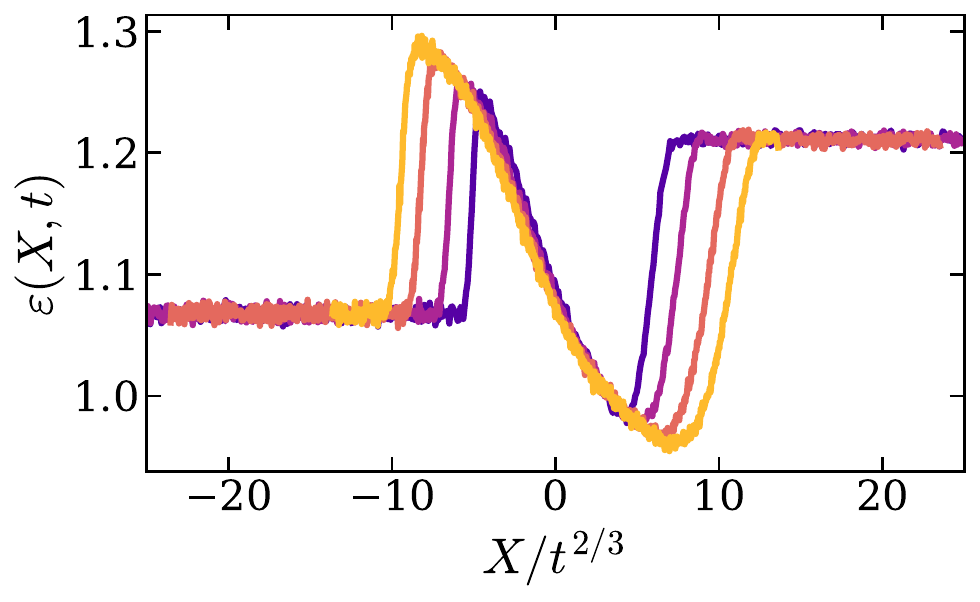}
\includegraphics[width=0.24\textwidth]{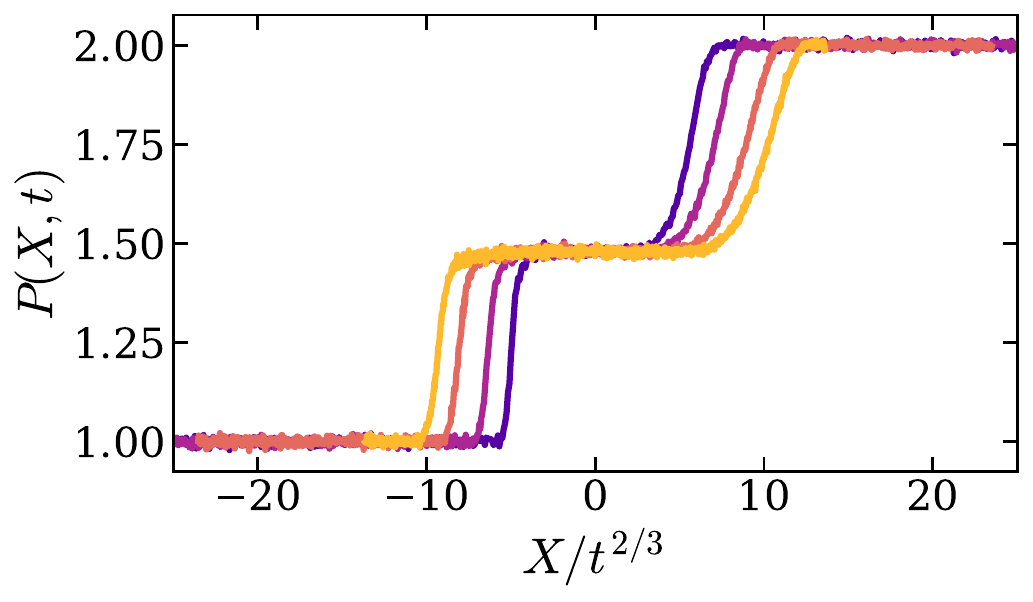}
\caption{
Comparison of the Euler solution and microscopic simulations for an unequal-pressure domain wall. The initial states are \(\cf_L=(r_L,0, \e_L)\) and \(\cf_R=(r_R,0, \e_R)\), with \(P_L \neq P_R\). The top panel shows the unscaled plots while the middle panel shows the same data with ballistic scaling. In the middle panel, the colored lines show ensemble averaged microscopic data plotted against the similarity variable \(X/t\), while the solid lines show the Euler prediction obtained from the Riemann solution construction. The lower panel shows $X/t^{2/3}$ scaling and we see somewhat better data collapse near the centre. The plots here are constructed after averaging $5\times 10^4$ samples with $\beta_L=1$, $\beta_R=2$, $P_L=1$ and $P_R=2$.
}
\label{fig:uneq_pressure_dw}
\end{figure}

Fig. \ref{fig:uneq_pressure_dw} shows the evolution of the hydrodynamic fields with time and the comparison with Euler solution shows that the locations of the main hydrodynamic structures in the microscopic profiles are well captured by the Euler theory, and the observed profile clearly exhibits the three-wave structure predicted by the Riemann construction: a left acoustic wave, a stationary central discontinuity, and a right acoustic wave. Thus, at the level of the large-scale wave pattern, the unequal-pressure domain wall behaves as expected from Euler hydrodynamics.

At the same time, the numerical data also show deviations from the ideal Euler profile near the central region. These corrections are expected, since the Euler theory describes only the leading ballistic structure and neglects dissipative and fluctuating effects. We see that the central profiles show a somewhat better collapse when rescaled as $X/t^{2/3}$ which is characteristic of  KPZ scaling. Such a scaling is consistent with the predictions of nonlinear fluctuating hydrodynamics for one-dimensional anharmonic chains, in which the two sound modes are in the KPZ universality class and the heat mode is super-diffusive \cite{spohn2014NFHT}.  Hence, this broadening can possibly be explained by taking into account diffusive and noise terms in the hydrodynamics, but performing this analysis in the present far-from-equilibrium situation is a challenging problem.

\subsubsection{\bf Equal-pressure domain wall}
\label{sec:eq-pressure}
We now consider a second domain wall initial condition in which the two sides
have different temperatures but the same pressure. We take $P_L = P_R = 0$. 
This case is qualitatively different from the unequal-pressure domain wall
discussed above. Since the two sides have the same pressure and zero velocity,
the conditions Eq.~\eqref{eq: RH_dw} for a stationary contact discontinuity are already satisfied. Thus, the initial
jump is simply a stationary contact discontinuity, and at the Euler level, the profile does not evolve in time.
\begin{figure}[t]
\centering
\includegraphics[width=0.24\textwidth]{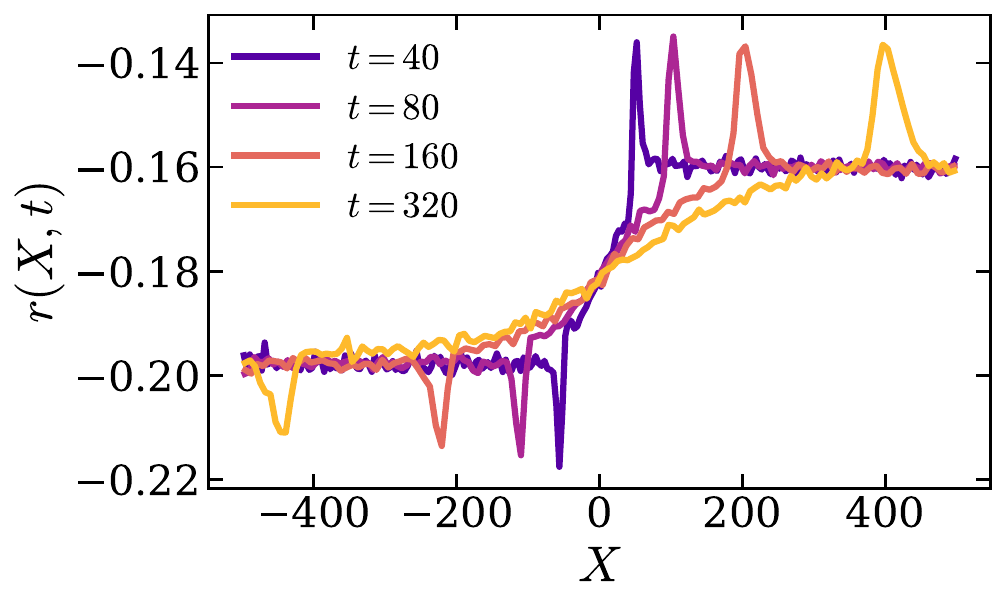}
\includegraphics[width=0.24\textwidth]{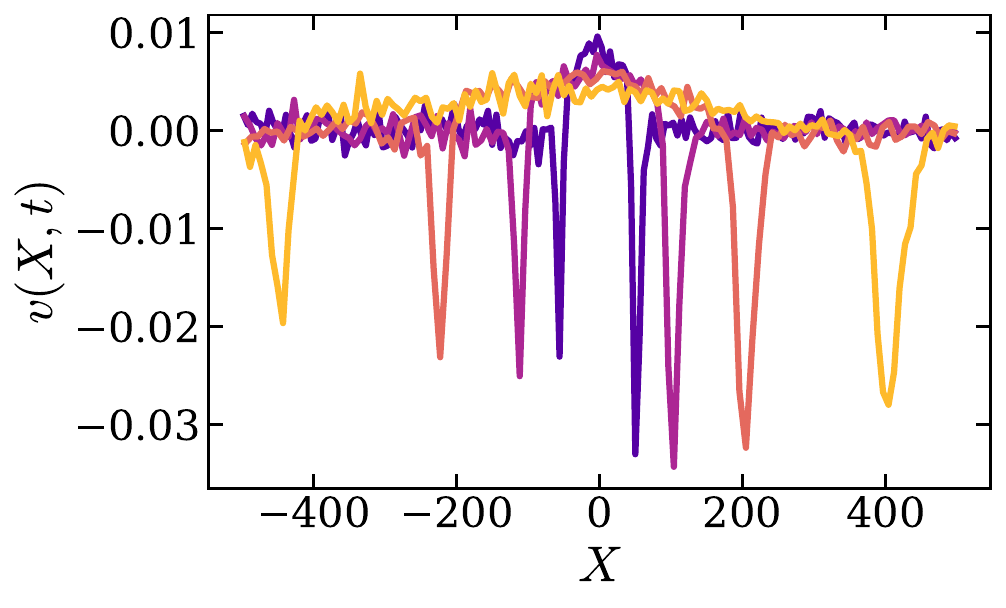}
\includegraphics[width=0.24\textwidth]{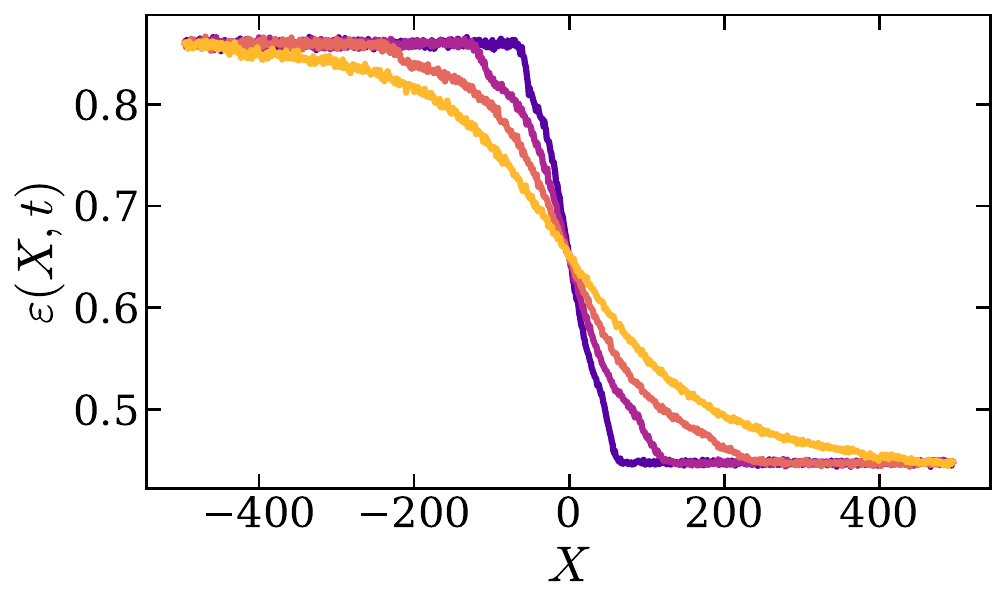}
\includegraphics[width=0.24\textwidth]{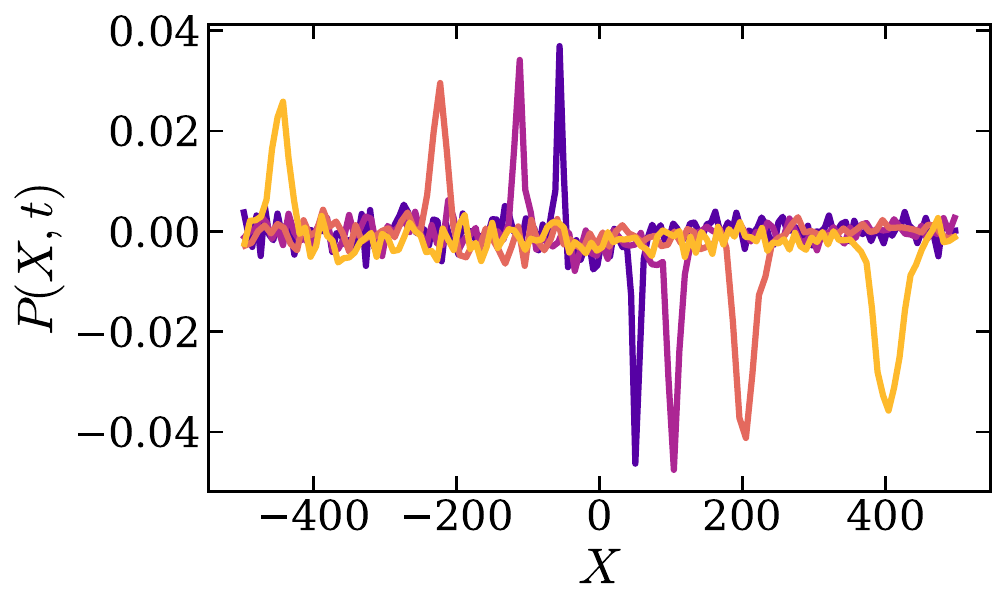}
\includegraphics[width=0.24\textwidth]{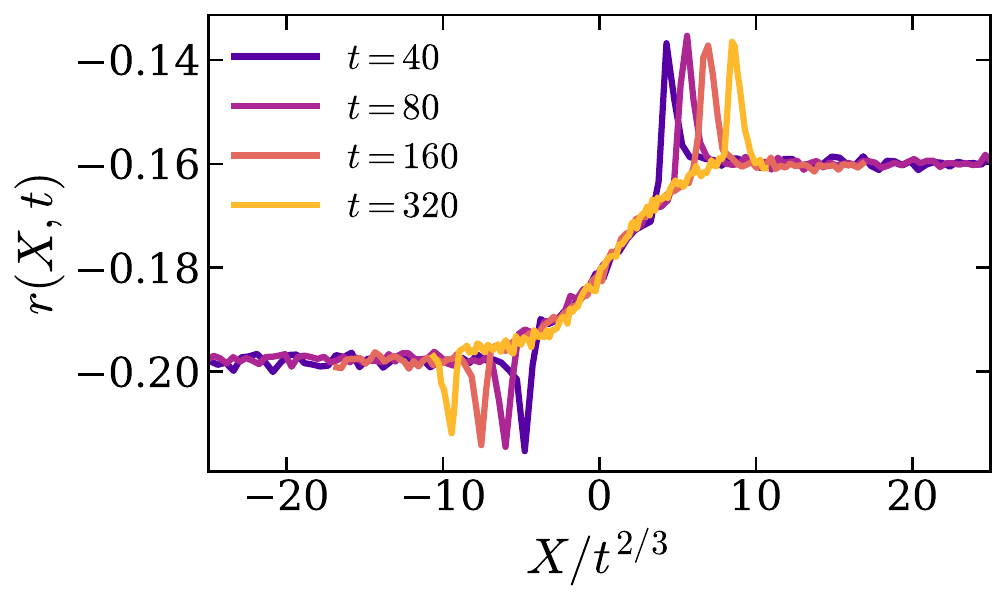}
\includegraphics[width=0.24\textwidth]{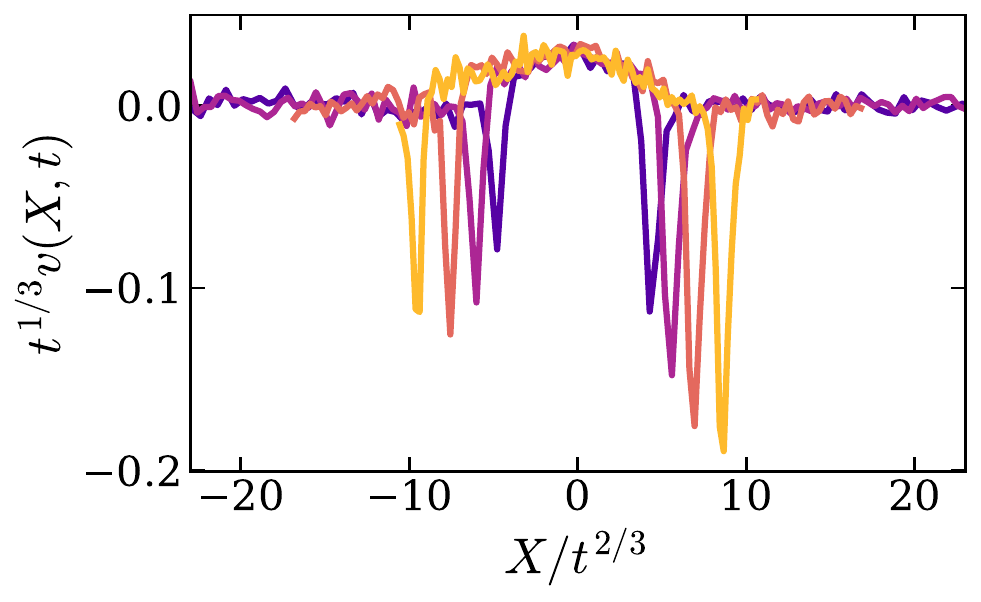}
\includegraphics[width=0.24\textwidth]{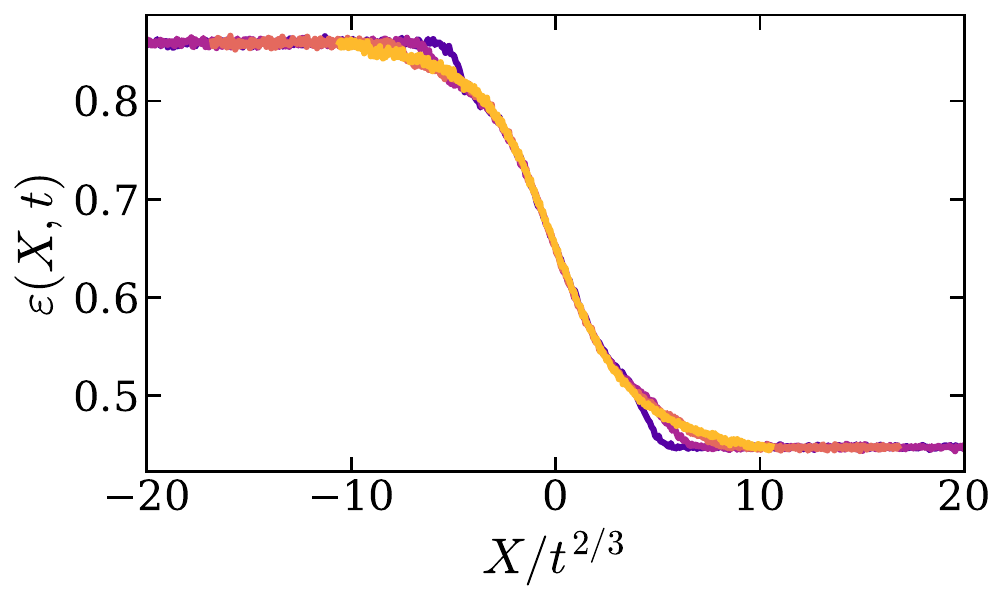}
\includegraphics[width=0.24\textwidth]{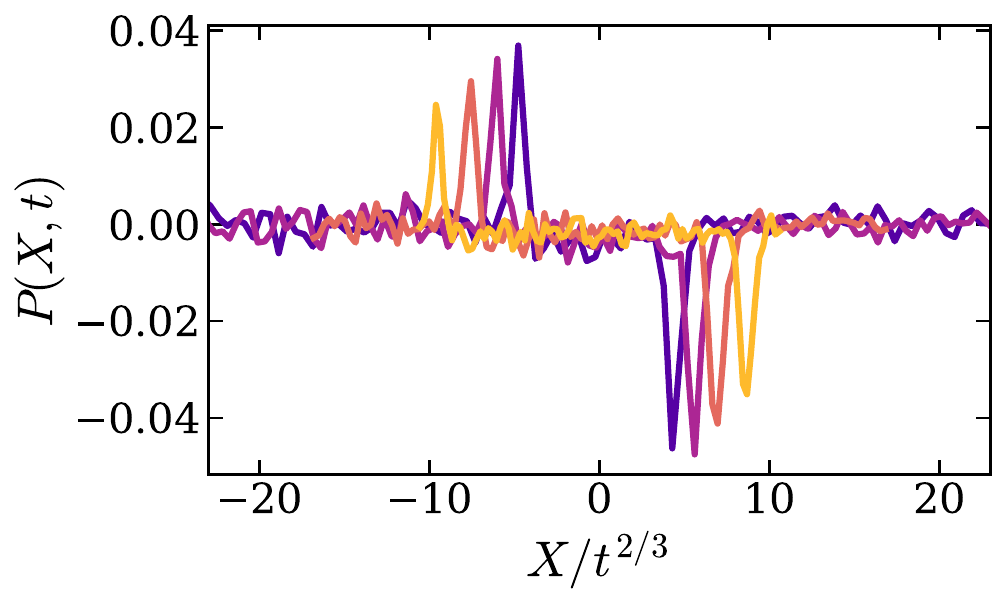}
\includegraphics[width=0.24\textwidth]{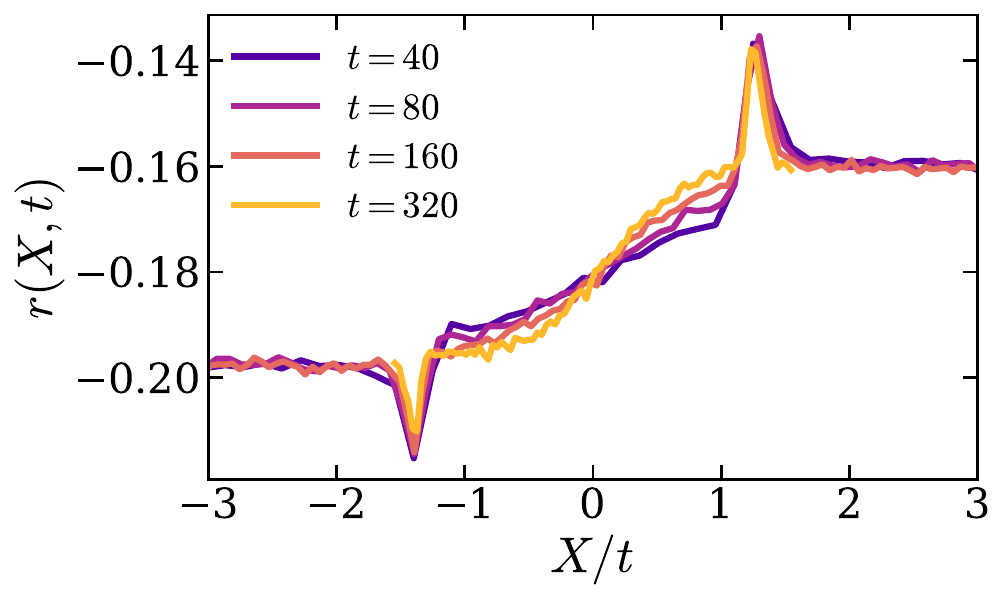}
\includegraphics[width=0.24\textwidth]{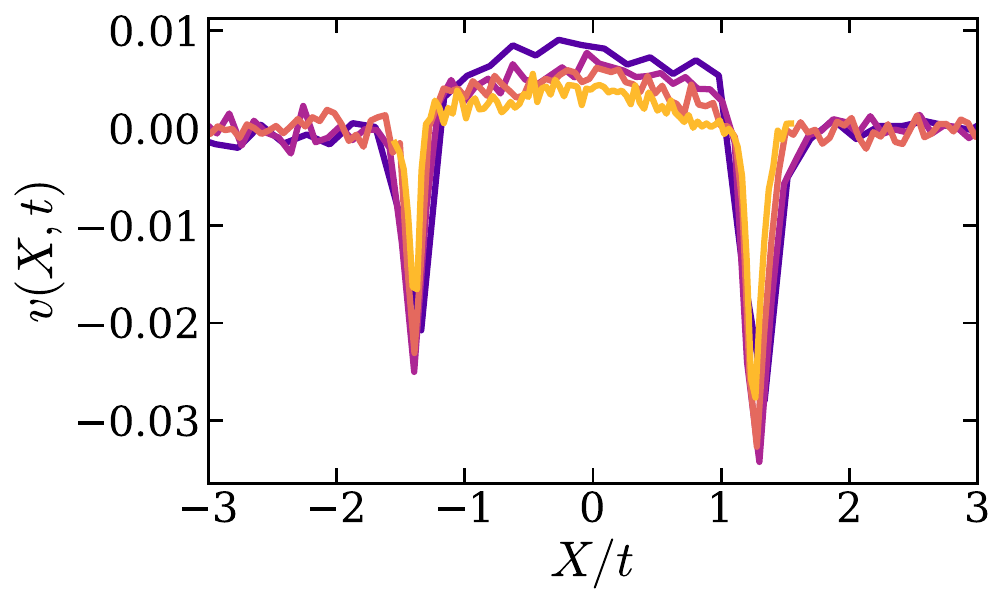}
\includegraphics[width=0.24\textwidth]{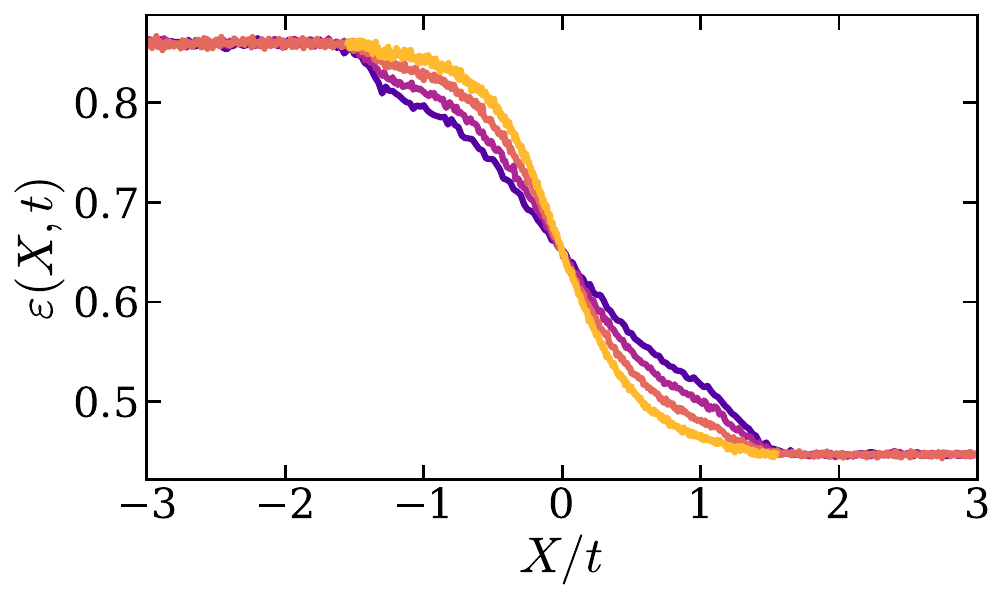}
\includegraphics[width=0.24\textwidth]{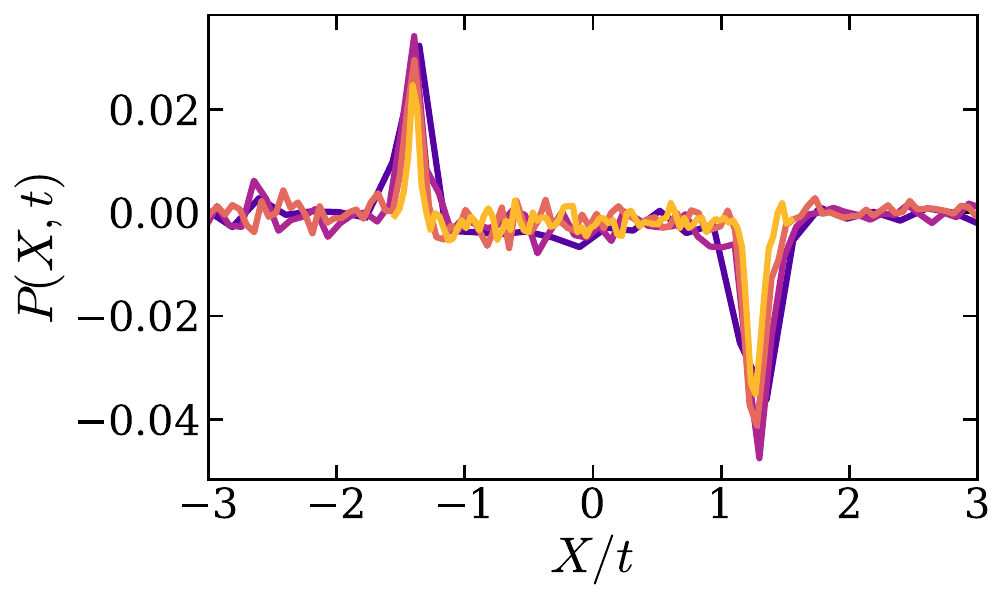}
\caption{
Evolution of an equal-pressure domain wall with $\beta_L=1$, $\beta_R=2$, $P=0$ and $v=0$, with $10^5$ samples averaged. The top panel shows the unscaled plots and the middle panel shows that the best collapse for the central regions is by $X/t^{2/3}$. There are moving spikes in the profiles that scale ballistically as shown in the bottom panel.
}
\label{fig:eq_pressure_dw}
\end{figure}
In the microscopic simulation, however, the initial discontinuity at $X=0$ does evolve and the microscopic profiles broaden, as shown in Fig. \ref{fig:eq_pressure_dw}. Since the Euler equations admit no such evolution for this initial data, the dynamics seen is entirely due to beyond-Euler corrections, i.e., to dissipative and fluctuating terms in the hydrodynamic description. 

In this case we find that the profiles show good collapse in the central region for the following scaling forms
\begin{align}
r(X,t) \simeq F\!\left(\frac{X}{t^{2/3}}\right),~~ \vf(X,t) \simeq \f{1}{t^{1/3}}\, G\!\left(\frac{X}{t^{2/3}}\right),~~ \varepsilon(X,t) \simeq H\left(\frac{X}{t^{2/3}}\right).
\end{align}
We observe again the scaling exponent $2/3$ that appeared in the central region of the unequal-pressure case in Sec.~\ref{sec:uneq_pressure}. In the equal-pressure setup the Euler contribution is absent, so the entire observed evolution is the fluctuating-hydrodynamic response to the initial jump. The collapse in Fig.~\ref{fig:eq_pressure_dw} thus provides a particularly clean test of the universal scaling, free from the moving acoustic shocks and rarefactions that complicated the unequal-pressure case.

Apart from the central region that exhibits KPZ scaling, we notice, in Fig.~\ref{fig:eq_pressure_dw}, sharp spikes in all the field profiles. These evolve ballistically with speeds that match the sound speeds on the two sides. At the moment, we do not have a detailed explanation for the spikes.

\subsection{\textbf{Blast Problem}}
\label{subsec:bp}
We now consider a blast-type nonequilibrium initial condition for the FPUT chain. The system is prepared in a state consisting of a localized energetic core embedded inside a cold ambient background. More precisely, the chain is initialized such that a finite region near the origin is sampled from an equilibrium state with parameters $(\beta_c,P_c)$, while the ambient region is at temperature $T_a=0$ with pressure $P_a$. The full chain is then evolved under the Hamiltonian dynamics and we ask how the central disturbance propagates into the ambient medium at long times. Previous studies in hard particle systems \cite{chakraborti2021,ganapa2021blast} have reported the appearance of self-similar TvNS blast waves in suitable parameter regimes and detailed verifications were obtained from molecular dynamic simulations of the scaling solution of the Euler equations. Motivated by these observations, we investigate whether analogous scaling arises in the FPUT chain.

\subsubsection{\bf The TvNS blast wave solution: necessary conditions}
A classical framework for understanding such problems is provided by the Taylor-von Neumann-Sedov theory of blast waves~\cite{taylor1950formation,vonneumann1947blast,sedov1946}. In this picture, the blast evolves into a self-similar expanding shock wave whose position is determined by dimensional analysis. In the case of a generic gas, let $E$ denote the total injected energy and let $\rho_a$ denote the ambient mass density. In one-dimension, the energy contained in a region of size $R$ is estimated as $E \sim \rho_a R \dot R^2 $. This gives $R(t) \sim \left(\frac{E}{\rho_a}\right)^{1/3} t^{2/3}$. Here $R(t)$ denotes the position of the propagating front. The physical origin of this scaling is the fact that the only relevant scales are the conserved total energy, finite ambient density and time, which together determine the growth of the expanding blast region. A requirement for this is that the ambient pressure $P_a=0$. 

The requirement of zero ambient pressure can  be understood using the Rankine-Hugoniot conditions for a shock in the following way: we define the scaling variable $\zeta=\frac{x}{R(t)}$ and assume the self-similar forms $\rho_-(x,t)=\rho_a+f(\zeta)$, $u_-(x,t)=\frac{x}{t}g(\zeta)$, $P_-(x,t)=\left(\frac{x}{t}\right)^2h(\zeta)$. 
%and use the ideal gas equation of state $P_-(x,t)=\rho_-(x,t)T_-(x,t)$. 
The state ahead of the shock is the ambient state $\rho_+=\rho_a$, $u_+=0$, $P_+=P_a$. The shock is located at a fixed value of the similarity variable, which is $\zeta=1$. Thus, immediately behind the shock, $\rho_-=\rho_a+f(1)$, $u_-=\frac{R(t)}{t}g(1)$, and $P_-=\left(\frac{R(t)}{t}\right)^2h(1).$ The mass Rankine-Hugoniot condition is $\dot R[\rho]=[\rho \vf]$. Substituting the scaling forms and taking $R(t)\sim t^\alpha,$ the mass jump condition becomes 
\begin{equation}
\alpha f(1)
=
\left(\rho_a+f(1)\right)g(1)
\end{equation}
with no explicit time dependence. Now consider the momentum Rankine-Hugoniot condition $\dot R[\rho \vf]=[\rho \vf^2+P]$. Substituting the self-similar forms gives
\begin{equation}
\alpha\left(\rho_a+f(1)\right)g(1)
=
\left(\rho_a+f(1)\right)g(1)^2
+
h(1)
-
P_a t^{2-2\alpha}.
\end{equation}
Thus, for any non-ballistic scaling $\alpha\neq 1$, the momentum jump condition contains explicit time dependence unless $P_a=0$.

Hence, to observe a TvNS type blast shock wave, the ambient pressure has to be $0$. However, this is not sufficient. We also need the ambient sound speed $c_a$ to be $0$. To see this, consider, at late times, the weak shock limit where the jump across the shock becomes small, $\cf_b = \cf_a + \delta \cf$, and linearizing the Rankine-Hugoniot conditions around the ambient state gives $\dot R\delta \cf = A(\cf_a)\,\delta \cf$. Thus, in the weak shock limit, the shock speed approaches one of the characteristic velocities of the Euler equations, $\dot R \to \lambda_i(\cf_a)=c_a$. However, for the one-dimensional blast, $R(t)\sim t^{2/3}$, and the front velocity scales as 
$\dot R(t)\sim t^{-1/3}\to 0$ at long times. Consistency with the weak shock limit therefore requires $c_a = 0.$ We are led to the conclusion that asymptotic TvNS scaling requires both $P_a = 0$, and $c_a = 0$.

\subsubsection{\bf Obstruction to TvNS scaling in FPUT}
We now investigate whether these conditions along with $r_a \neq 0$ (since $\rho_a = \frac{1}{r_a}$ needs to be finite) can be realized in the FPUT chain at zero temperature.

For the quartic FPUT potential $V(r) = \frac{k_2}{2}r^2 + \frac{k_3}{3}r^3 + \frac{k_4}{4}r^4$, the equilibrium ambient state at $T=0$ is determined by the global minimum of the effective potential. For vanishing pressure this reduces to the condition $V'(r_a)=0$. Also, at zero temperature, the sound speed is determined by the curvature of the potential at the ambient state, $c_a^2 = V''(r_a)$ (from Eq.(\ref{eq:DW-sound-speed})). We therefore want to find a finite ambient stretch $r_a$ satisfying both $V'(r_a)=0$, and $V''(r_a)=0$. Solving $V'(r)=k_2 r + k_3 r^2 + k_4 r^3=0$, and $V''(r)=k_2 + 2k_3 r + 3k_4 r^2=0$, we get, for $r_a \neq 0$, $r_a = -\frac{k_3}{2k_4}$, and $k_2 = \frac{k_3^2}{4k_4}$. For these values we find, $V'(r)=k_4 r(r-r_a)^2$, which implies that $V'(r)$ does not change sign across $r=r_a$. Thus $r_a$ is not a local minimum of the potential, but rather a stationary inflection point. Since the zero temperature equilibrium state must correspond to a global minimum of the potential, such a point cannot describe a physical ambient state. Consequently, for the FPUT chain at zero temperature the TvNS blast wave argument fails. 

In the next section, we show  microscopic simulations of the blast problem in the FPUT system for different initial conditions that support this absence of TvNS scaling.

\subsubsection{\bf Microscopic simulations}
In our microscopic simulations, we study three representative cases, based on which of the conditions on the ambient state are satisfied. In all these cases, we take the ambient temperature, $T_a=0$ and a small number of sites at the centre are initiated at a finite temperature and pressure. In all cases we do not see a sharp shock and the front of the blast propagates ballistically. Below, we discuss other details.

\begin{figure}[H]
\centering
\includegraphics[width=0.329\textwidth]{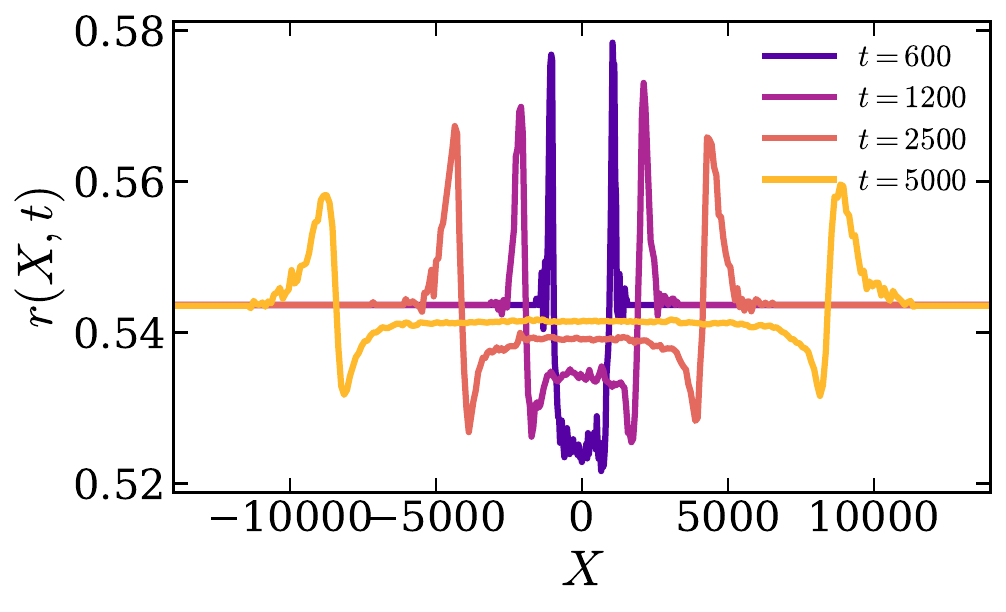}
\includegraphics[width=0.329\textwidth]{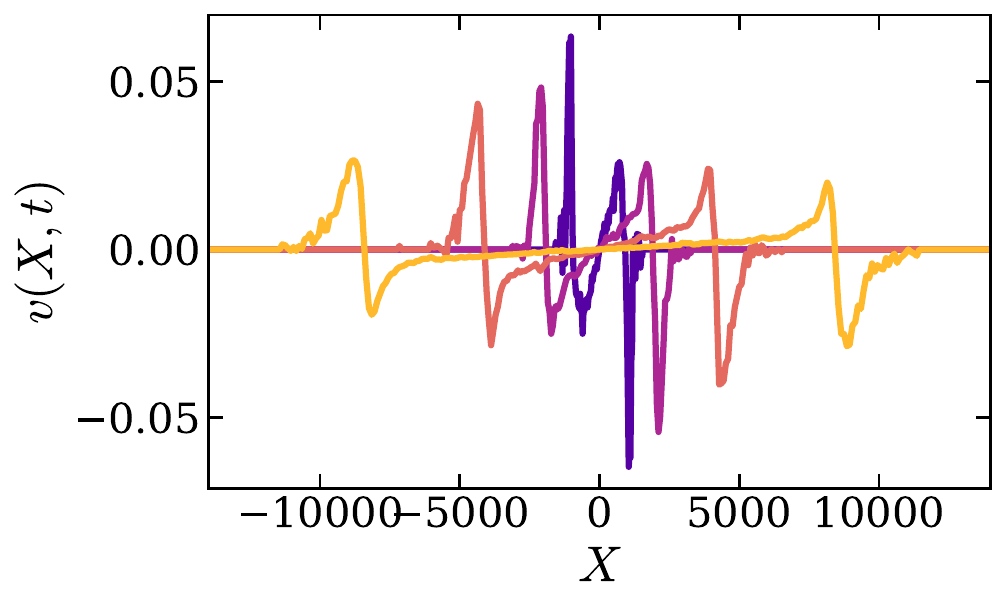}
\includegraphics[width=0.329\textwidth]{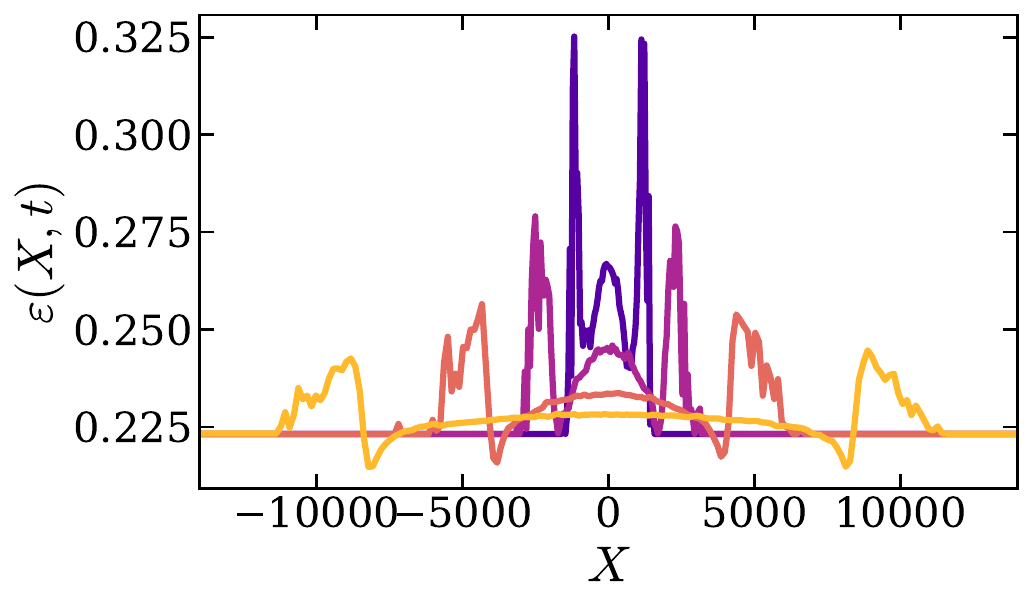}\\
\includegraphics[width=0.329\textwidth]{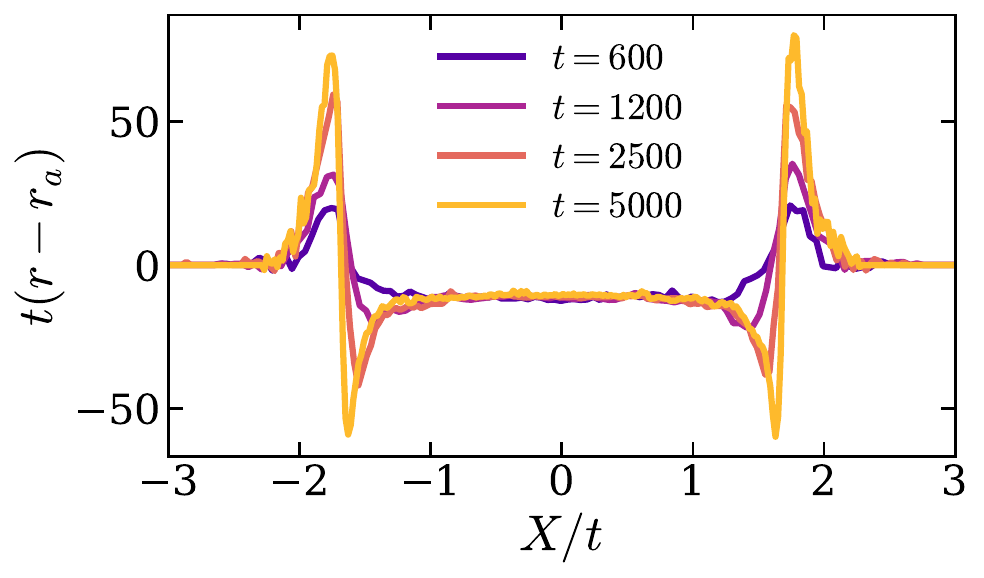}
\includegraphics[width=0.329\textwidth]{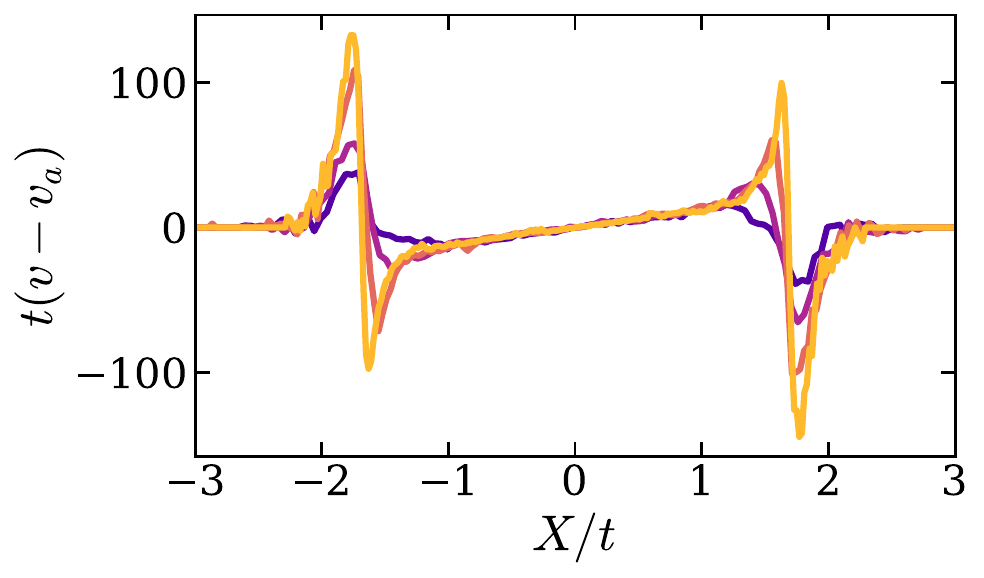}
\includegraphics[width=0.329\textwidth]{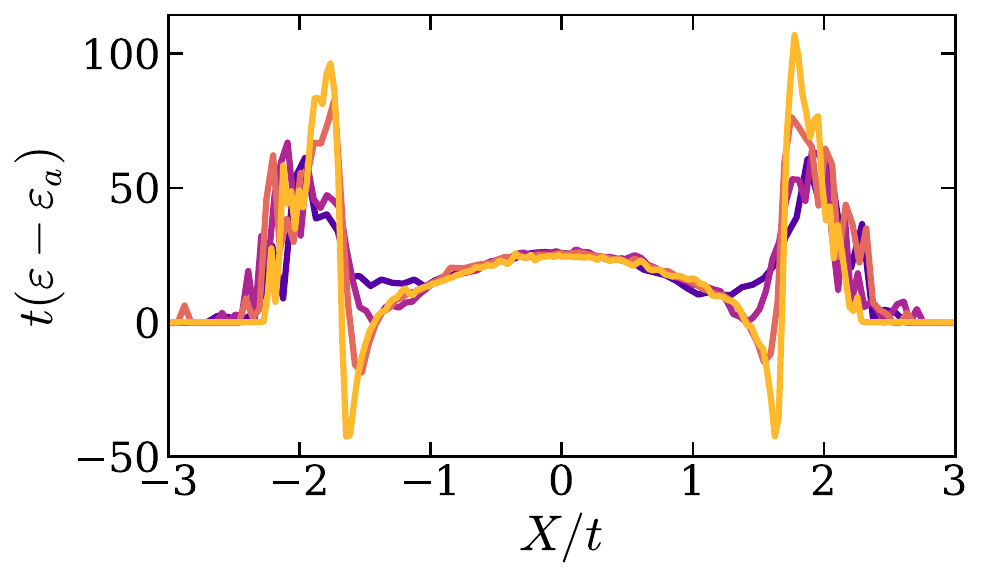}
\caption{
Case A: Blast initial condition with $P_a \neq 0$, $r_a \neq 0$, $c_a \neq 0$ and $(k_2, k_3, k_4)=(1,1,1)$. The top panel shows the time evolution of the profiles while the bottom panel shows results after a ballistic collapse.
}
\label{ambient_r_blast}
\end{figure}

{\bf Case A}: In Fig. \ref{ambient_r_blast}, we plot the unscaled hydrodynamic profiles for the three fields. Under a ballistic rescaling, i.e., $\phi(X,t) \;=\; \frac{1}{t}\, F_\phi\!\left(\frac{X}{t}\right),\quad \phi \in \{\, r-r_a,\; \vf-\vf_a,\; \varepsilon-\varepsilon_a\,\},$ we see that the front positions collapse onto fixed values of $X/t$, confirming that they move ballistically. The wavefronts themselves, however, do not collapse well. The central region between the two wavefronts does collapse well for all three
fields, indicating that the bulk of the profiles behaves as a ballistically spreading background with conserved area.

\begin{figure}[H]
\centering
\includegraphics[width=0.329\textwidth]{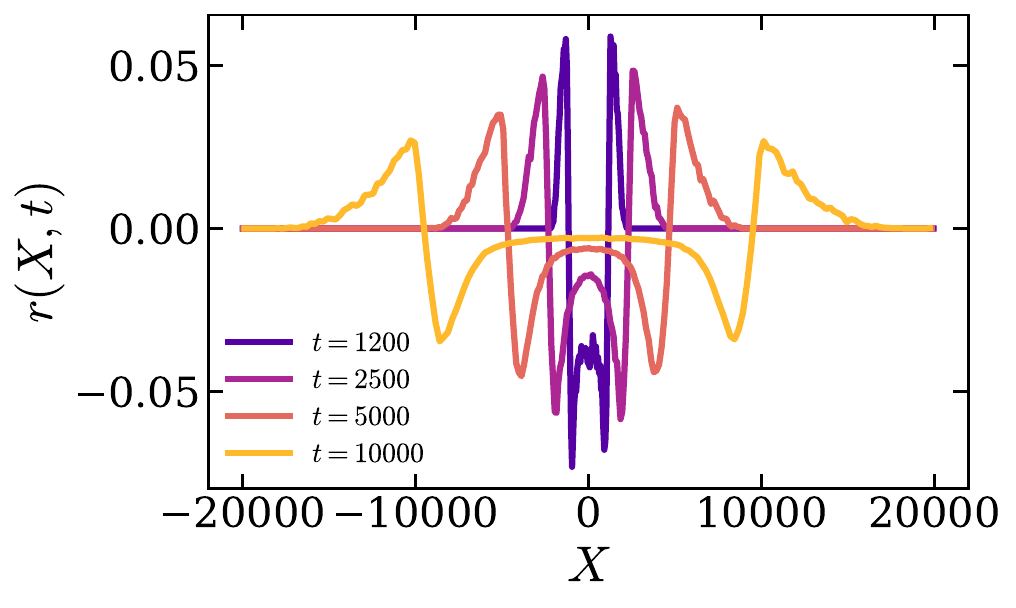}
\includegraphics[width=0.329\textwidth]{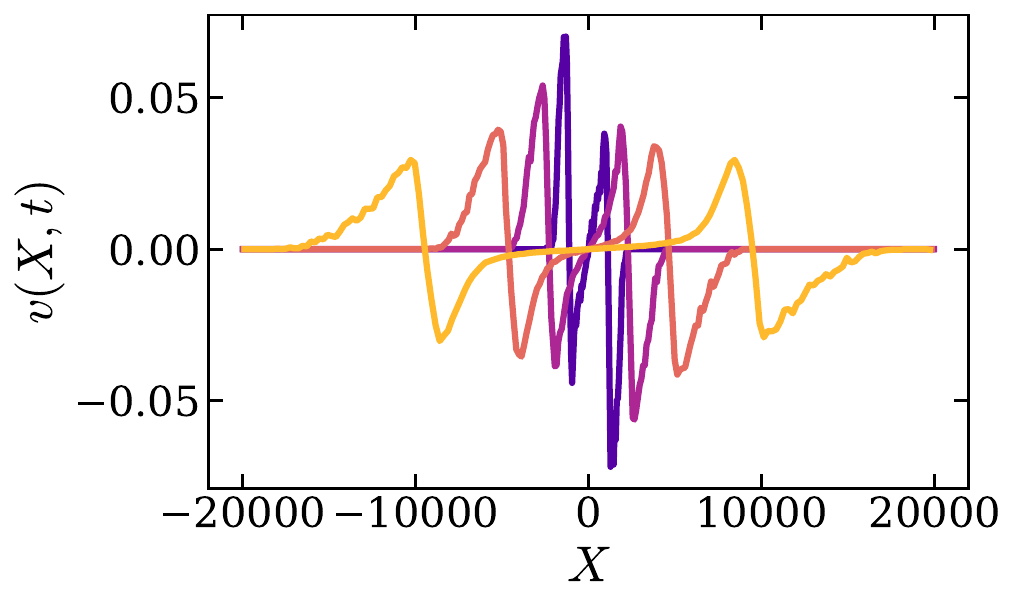}
\includegraphics[width=0.329\textwidth]{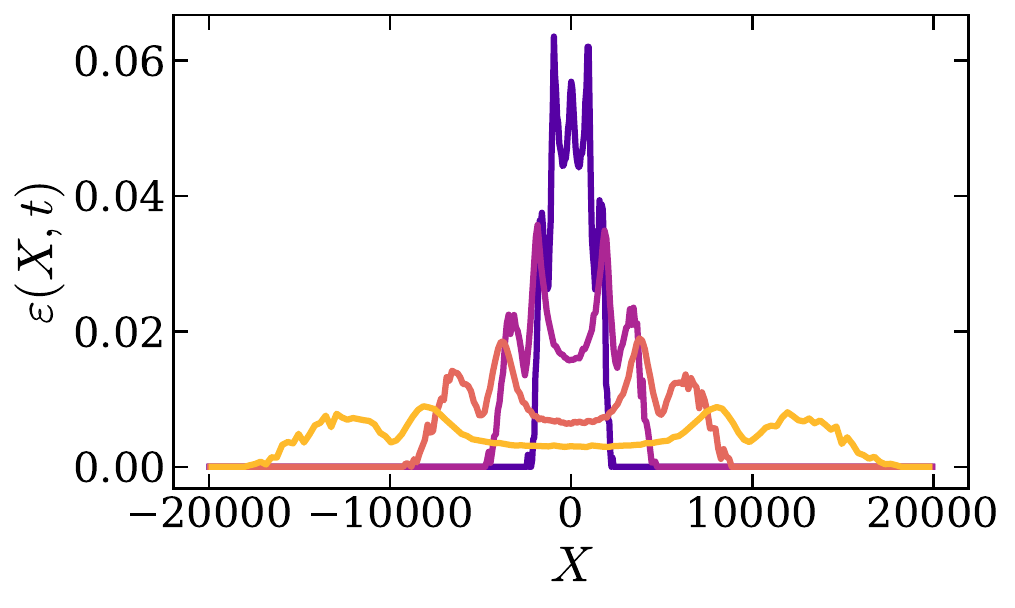}\\
\includegraphics[width=0.329\textwidth]{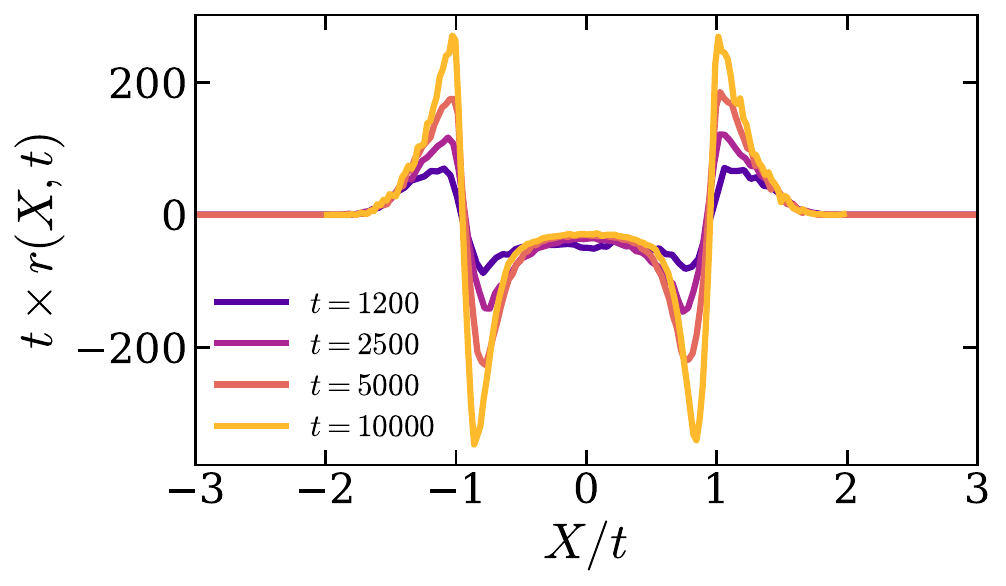}
\includegraphics[width=0.329\textwidth]{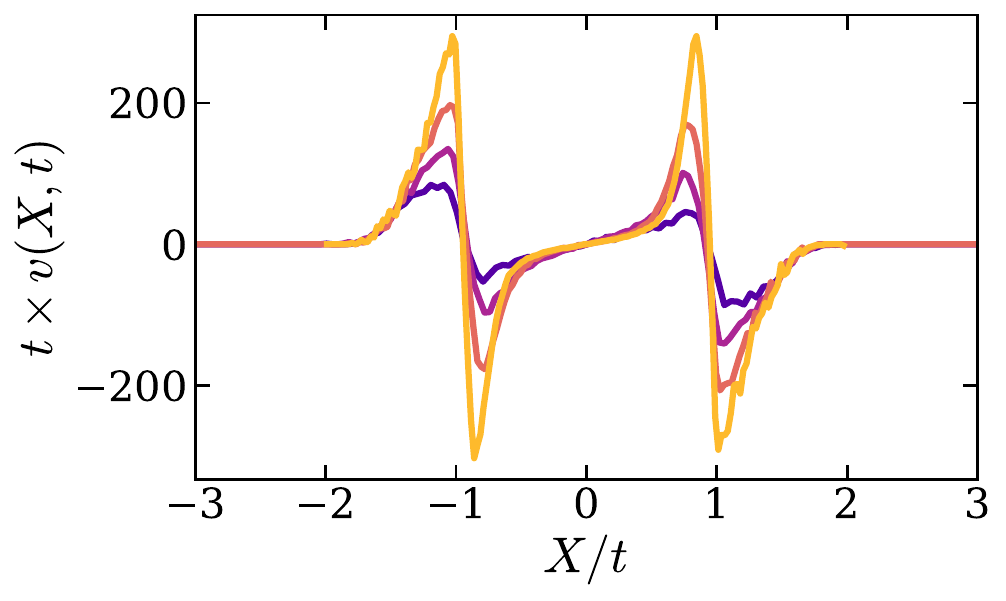}
\includegraphics[width=0.329\textwidth]{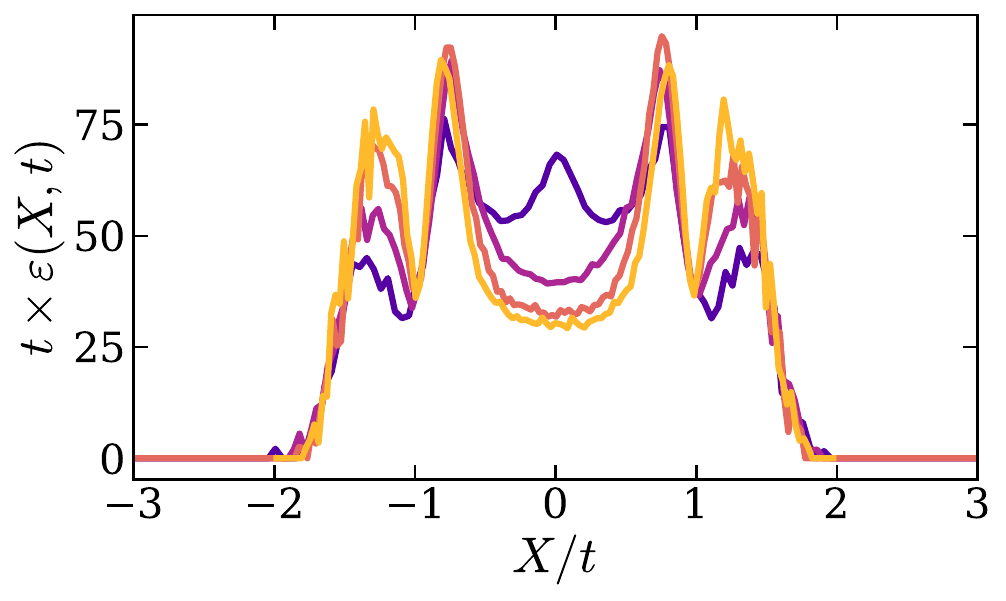}
\caption{
Case B: Blast with $P_a = 0$, $r_a = 0$, $c_a \neq 0$ and $(k_2,k_3,k_4)=(1,1,1)$. The top panel shows the time evolution of the unscaled profiles while the bottom panel shows profiles after ballistic scaling.
}
\label{fig:40k_blast}
\end{figure}

{\bf Case B}: In Fig. \ref{fig:40k_blast}, we observe outward-propagating pulses whose front positions are again ballistic. The structure is, however, qualitatively different from Case~A: neither the wavefronts nor the central region collapse under the ballistic rescaling.

\begin{figure}[H]
\centering
\includegraphics[width=0.329\textwidth]{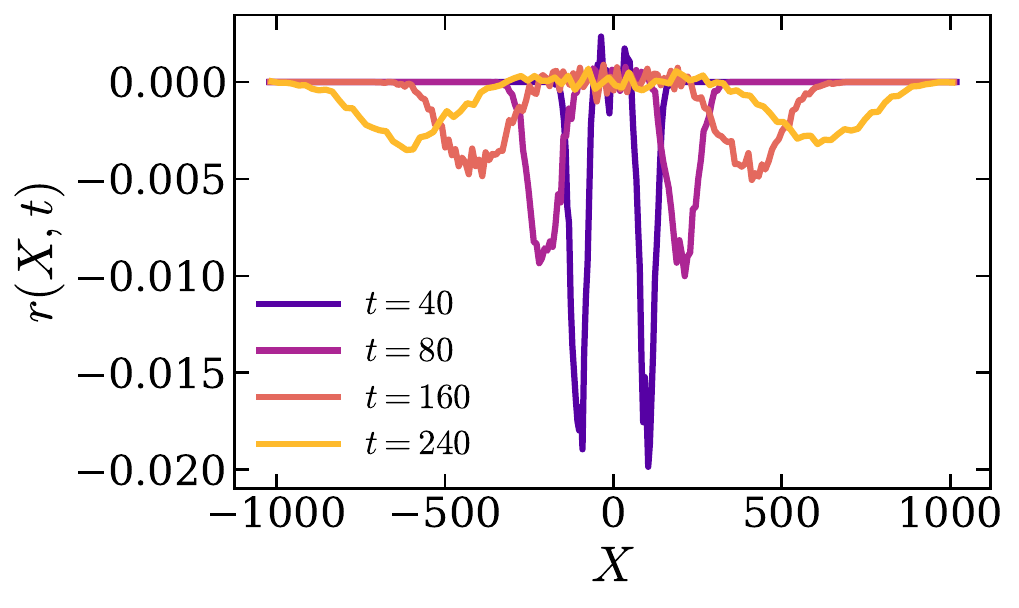}
\includegraphics[width=0.329\textwidth]{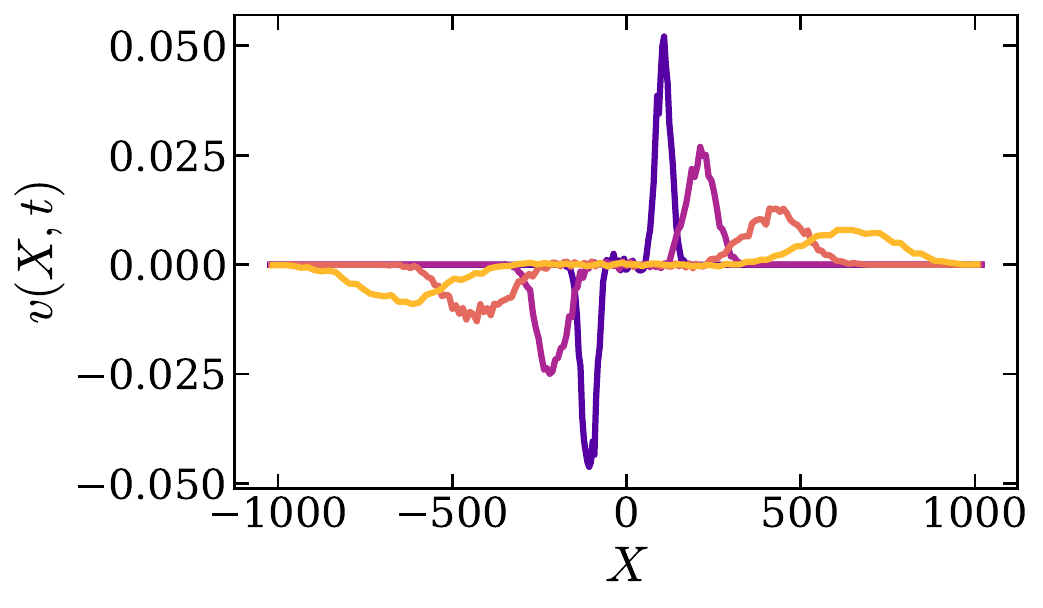}
\includegraphics[width=0.329\textwidth]{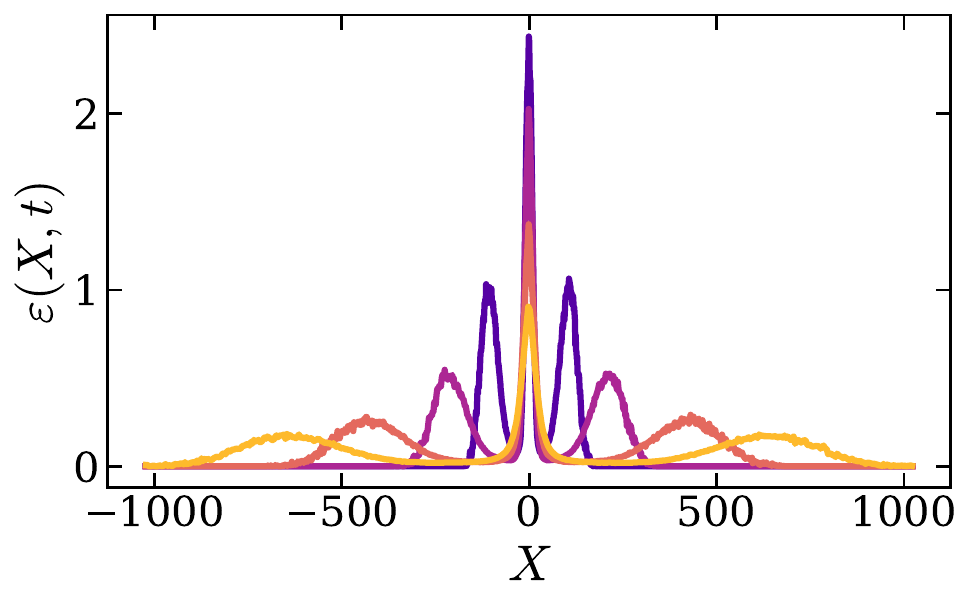}\\
\includegraphics[width=0.32\textwidth]{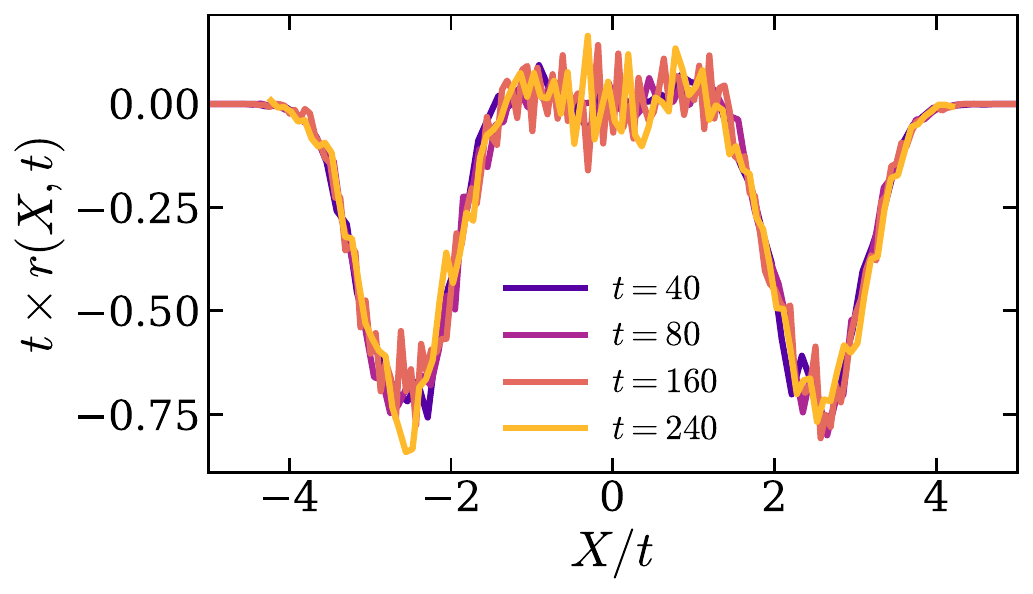}
\includegraphics[width=0.32\textwidth]{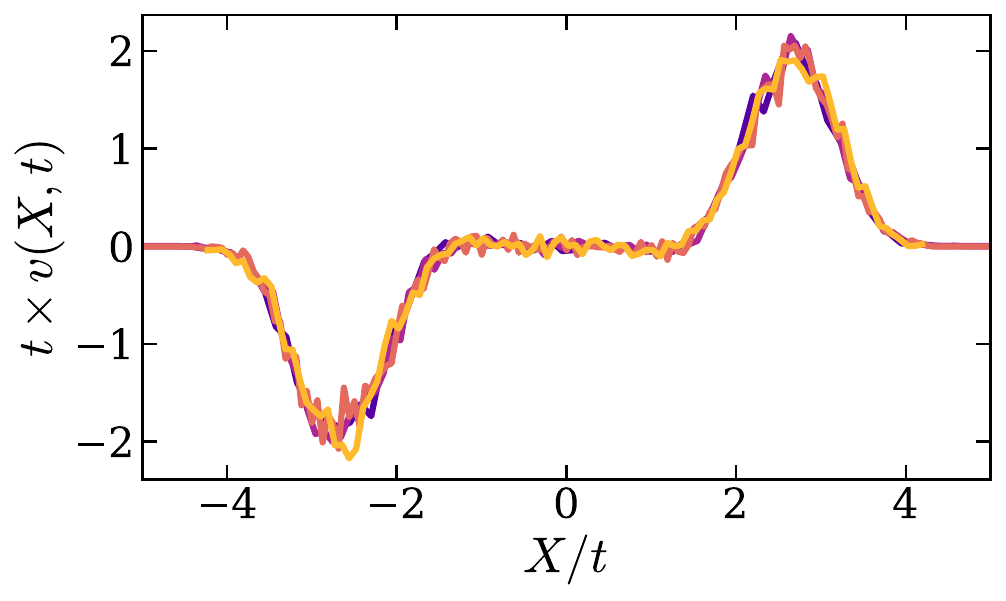}
\includegraphics[width=0.32\textwidth]{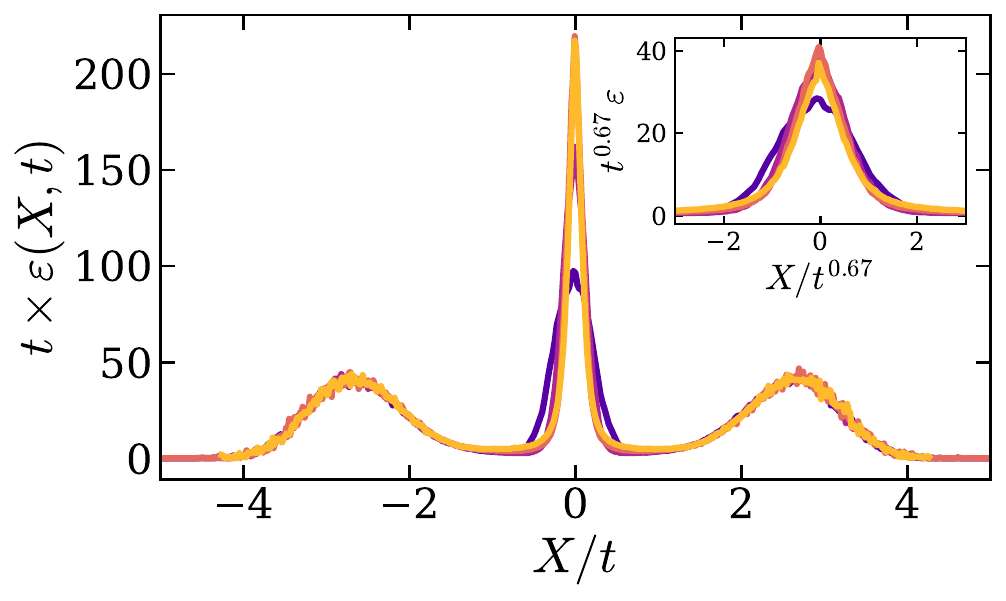}
\caption{Case C: Blast with $P_a = 0$, $r_a = 0$, $c_a = 0$ and $(k_2,k_3,k_4)=(0,0,1)$. The top panel shows the unscaled profiles and the bottom panel shows a ballistic scaling of the profiles. The inset in the scaled $\varepsilon$ plot shows a $X/t^{2/3}$ scaling collapse of the central $\varepsilon$ peak.
}
\label{fig: c_0_blast}
\end{figure}

{\bf Case C}: Here we see a much more structured three-wave pattern in Fig.~\ref{fig: c_0_blast}: two outward-propagating pulses, present in all three fields, together with an additional component that remains centred at $X = 0$, and that grows, in the $\varepsilon$ profile.  Under the ballistic rescaling, the two outward pulses in $r$, $v$ and $\varepsilon$ collapse cleanly. The central peak in $\varepsilon$, on the other hand, does not collapse; it seems to spread super-diffusively with an exponent $2/3$, as shown in the inset of Fig. \ref{fig: c_0_blast}.

% \begin{figure}[H]
% \centering
% \includegraphics[width=0.329\textwidth]{Images/blast_new/r_single.pdf}
% \includegraphics[width=0.329\textwidth]{Images/blast_new/v_single.pdf}
% \includegraphics[width=0.329\textwidth]{Images/blast_new/eps_single.pdf}
% \caption{
% Comparison of two coarse-grained single-realization profiles with an ensemble-averaged profile in Case A for $t=5000$. 
% }
% \label{single}
% \end{figure}

All the profiles presented in this section were obtained by averaging over a number of independent realizations of the initial state. In Cases A and B, we average $100$ samples on a chain of size $N=4\times10^4$ and coarse-grain to get the final plots whereas in Case C, we average over $5\times 10^4$ samples on a chain of size $N=2048$. 

%In Fig.~\ref{single} we compare two coarse-grained individual realizations of the three fields with the corresponding ensemble-averaged profile at a fixed time $t$, for Case A. The single-realization profiles seem to differ from each other and from the average. The mismatch is largest near the propagating fronts but also significant in the central region. Thus, at the system sizes and times accessible to us, a single realization does not reproduce the deterministic hydrodynamic profile, which raises the question of whether these fields self-average.

% \paragraph{$P_a=0,$ $r_a=0,$ $c_a\neq 0$ (II)}

% \begin{figure}[H]
% \centering
% \includegraphics[width=0.32\textwidth]{Images/temp/p_10_T_10_ell_profile.png}
% \includegraphics[width=0.32\textwidth]{Images/temp/p_10_T_10_u_profile.png}
% \includegraphics[width=0.32\textwidth]{Images/temp/image.png}
% \caption{
% \ad{temp}
% }
% \label{}
% \end{figure}

\section{Summary and Discussions}
\label{sec: summary}
A large number of studies have focused on the question of thermalization of the FPUT chain, starting from special initial conditions. There have also been many studies on the form of space-time correlations of the conserved fields in FPUT chains prepared in equilibrium states, where anomalous scaling has been observed. The theory of nonlinear fluctuating hydrodynamics has led to some theoretical understanding including the connections to KPZ physics.

A somewhat less explored question is the application of hydrodynamics to far-from-equilibrium phenomena. Here we studied two well known nonequilibrium initial states, the domain wall and the blast wave, which have been studied for other model systems but so far not applied to the FPUT system. In addition we presented heuristic derivations of the hydrodynamic equations, at the Euler and Navier-Stokes-Fourier level. 

For the domain wall, two different situations were probed, one with different pressures on the two sides and the other with equal pressures and it was found that they showed significant qualitative differences. 
The solution of the Euler equations to find the evolution of shocks and rarefaction waves is non-trivial since the equation of state of the FPUT system has to be evaluated numerically. For the case with different pressures, we observed shocks and rarefaction waves and found good agreement with the predictions from the Euler equations. For the equal pressure case, there is no evolution at the Euler scale. In this case we observe KPZ type scaling of the field profiles--- we believe this can be explained by inclusion of  the dissipative and noise terms in the hydrodynamic equation.   

For the problem of a blast in a cold gas, we argued that the FPUT system cannot show the TvNS type scaling solutions that have been observed for the one-dimensional hard point gas. Our simulations support this. For all the three different choices of parameters that were studied, we did not see shocks  and the front of the profile moves ballistically in all cases. This contrasts the sharp shock fronts growing as $t^{2/3}$ observed in the hard point gas~\cite{chakraborti2021}. For the choice $P_a=0,r_a=0,c_a=0$, we observed a good ballistic scaling of all the fields, with the energy field showing a $t^{2/3}$ scaling in the core region. 

A complete theoretical understanding of far-from-equilibrium dynamics of the FPUT system requires a careful consideration of the effects of dissipation and noise as is already known from studies of nonequilibrium heat conduction. This appears to be a challenging problem that requires further studies.

\section*{Acknowledgments}{We thank Anupam Kundu for valuable comments on the draft.  We  acknowledge support from the Department of Atomic Energy, Government of India, under Project No. RTI4001. A.D. acknowledges support from the J.C.~Bose Fellowship (JCB/2022/000014) of the Science and Engineering Research Board, Department of Science and Technology, Government of India.}

\section*{Declarations}
{\bf Data availability statement}: The numerical data presented in this work is available on request. \\

\noindent {\bf Conflicts of Interest}: The authors have no conflict of interest to disclose.

\bibliographystyle{unsrt}
\bibliography{references}

@article{olla2019,
    title={Role of conserved quantities in Fourier's law for diffusive mechanical systems},
    author={Olla, S.},
    journal={Comptes Rendus. Physique},
    volume={20},
    number={5},
    pages={429--441},
    year={2019}
}

@article{dhar2008,
    title={Heat transport in low-dimensional systems},
    author={Dhar, A.},
    journal={Advances in Physics},
    volume={57},
    number={5},
    pages={457--537},
    year={2008},
    publisher={Taylor \& Francis}
}

@article{spohn2014NFHT,
    title={Nonlinear fluctuating hydrodynamics for anharmonic chains},
    author={Spohn, H.},
    journal={J. Stat. Phys.},
    volume={154},
    number={5},
    pages={1191--1227},
    year={2014},
    publisher={Springer}
}

@article{lepri1997,
  title={Heat conduction in chains of nonlinear oscillators},
  author={Lepri, S. and Livi, R. and Politi, A.},
  journal={Phys. Rev. Lett.},
  volume={78},
  number={10},
  pages={1896},
  year={1997},
  publisher={APS}
}

@article{narayan2002,
  title={Anomalous heat conduction in one-dimensional momentum-conserving systems},
  author={Narayan, O. and Ramaswamy, S.},
  journal={Phys. Rev. Lett.},
  volume={89},
  number={20},
  pages={200601},
  year={2002},
  publisher={APS}
}

@article{van2012,
  title={Exact results for anomalous transport in one-dimensional Hamiltonian systems},
  author={Van Beijeren, H.},
  journal={Phys. Rev. Lett.},
  volume={108},
  number={18},
  pages={180601},
  year={2012},
  publisher={APS}
}

@article{mendl2013,
  title={Dynamic correlators of Fermi-Pasta-Ulam chains and nonlinear fluctuating hydrodynamics},
  author={Mendl, C. B. and Spohn, H.},
  journal={Phys. Rev. Lett.},
  volume={111},
  number={23},
  pages={230601},
  year={2013},
  publisher={APS}
}

@article{mai2007,
  title={Equilibration and universal heat conduction in fermi-pasta-ulam chains},
  author={Mai, T. and Dhar, A. and Narayan, O.},
  journal={Phys. Rev. Lett.},
  volume={98},
  number={18},
  pages={184301},
  year={2007},
  publisher={APS}
}

@article{saito2010,
  title={Heat conduction in a three dimensional anharmonic crystal},
  author={Saito, K. and Dhar, A.},
  journal={Phys. Rev. Lett.},
  volume={104},
  number={4},
  pages={040601},
  year={2010},
  publisher={APS}
}

@article{zhao2012,
  title = {Normal heat conduction in one-dimensional momentum conserving lattices with asymmetric interactions},
  author = {Zhong, Y. and Zhang, Y. and Wang, J. and Zhao, H.},
  journal = {Phys. Rev. E},
  volume = {85},
  issue = {6},
  pages = {060102(R)},
  numpages = {4},
  year = {2012},
  month = {Jun},
  publisher = {American Physical Society},
  doi = {10.1103/PhysRevE.85.060102},
  url = {https://link.aps.org/doi/10.1103/PhysRevE.85.060102}
}

@article{singh2023,
  title={Blast Waves in the Zero Temperature Hard Sphere Gas: Double Scaling Structure.},
  author={Singh, S. K. and Chakraborti, S. and Dhar, A. and Krapivsky, P. L.},
  journal={J. Stat. Phys.},
  volume={190},
  number={7},
  year={2023}
}

@article{chen2013,
  title={Diffusion of heat, energy, momentum, and mass in one-dimensional systems},
  author={Chen, S. and Zhang, Y. and Wang, J. and Zhao, H.},
  journal={Phys. Rev. E},
  volume={87},
  number={3},
  pages={032153},
  year={2013},
  publisher={APS}
}

@article{chen2016,
  title={Key role of asymmetric interactions in low-dimensional heat transport},
  author={Chen, S. and Zhang, Y. and Wang, J. and Zhao, H.},
  journal={Journal of Statistical Mechanics: Theory and Experiment},
  volume={2016},
  number={3},
  pages={033205},
  year={2016},
  publisher={IOP Publishing and SISSA}
}

@article{lepri2020,
  title={Too close to integrable: Crossover from normal to anomalous heat diffusion},
  author={Lepri, S. and Livi, R. and Politi, A.},
  journal={Phys. Rev. Lett.},
  volume={125},
  number={4},
  pages={040604},
  year={2020},
  publisher={APS}
}

@article{das2014b,
  title={Heat conduction in the $\alpha$- $\beta$ Fermi--Pasta--Ulam chain},
  author={Das, S. G. and Dhar, A. and Narayan, O.},
  journal={J. Stat. Phys.},
  volume={154},
  number={1},
  pages={204--213},
  year={2014},
  publisher={Springer}
}

@book{lepri2016,
  title={Thermal transport in low dimensions},
  author={Lepri, S. and Livi, R. and Politi, A.},
  volume={921},
  year={2016},
  publisher={Springer}
}

@article{durnin2021,
  title={Diffusive hydrodynamics of inhomogenous Hamiltonians},
  author={Durnin, J. and De Luca, A. and De Nardis, J. and Doyon, B.},
  journal={Journal of Physics A: Mathematical and Theoretical},
  volume={54},
  number={49},
  pages={494001},
  year={2021},
  publisher={IOP Publishing}
}

@misc{Fermi1955,
  title={Studies of the nonlinear problems},
  note={Los Alamos National Laboratory Report (LA1940), also in Collected Papers of Enrico Fermi 2},
  author={Fermi, E. and Pasta, J. and Ulam, S.},
  year={1955},
  publisher={Chicago: University of Chicago Press}
}

@article{Berman2005,
author = {Berman,G. P.  and Izrailev,F. M. },
title = {The {F}ermi-{P}asta-{U}lam problem: Fifty years of progress},
journal = {Chaos: An Interdisciplinary Journal of Nonlinear Science},
volume = {15},
number = {1},
pages = {15104},
year = {(2005)},
doi = {10.1063/1.1855036},
URL = { 
        https://doi.org/10.1063/1.1855036
    },
eprint = { 
        https://doi.org/10.1063/1.1855036    
}
}

@book{weissert1997,
  title={The genesis of simulation in dynamics: pursuing the Fermi-Pasta-Ulam problem},
  author={Weissert, T. P.},
  year={2012},
  publisher={Springer}
}

@ARTICLE{Benettin2013,
   author = {{Benettin}, G. and {Christodoulidi}, H. and {Ponno}, A.},
    title = "{The Fermi-Pasta-Ulam Problem and Its Underlying Integrable Dynamics}",
  journal = {J. Stat. Phys.},
     year = "(2013)",
    month = "Jul",
   volume = "152",
    pages = {195-212},
      doi = {10.1007/s10955-013-0760-6},
   adsurl = {https://ui.adsabs.harvard.edu/abs/2013JSP...152..195B}
}

@article{Ford1992,
author = "J. Ford",
title = "The {F}ermi-{P}asta-{U}lam problem: Paradox turns discovery",
journal = "Physics Reports",
volume = "213",
number = "5",
pages = "271 - 310",
year = "(1992)",
issn = "0370-1573",
doi = "https://doi.org/10.1016/0370-1573(92)90116-H",
url = "http://www.sciencedirect.com/science/article/pii/037015739290116H"
}

@InCollection{Gallavotti2008,
author="Gallavotti, G.",
title="Introduction to {F}{P}{U}",
editor="Gallavotti, Giovanni",
booktitle="The Fermi-Pasta-Ulam Problem: A Status Report",
year="(2008)",
publisher="Springer Berlin Heidelberg",
url="10.1007/978-3-540-72995-2_1"
}

@article{chakraborti2022,
  title={A splash in a one-dimensional cold gas},
  author={Chakraborti, S. and Dhar, A. and Krapivsky, P.},
  journal={SciPost Physics},
  volume={13},
  number={3},
  pages={074},
  year={2022}
}

@article{Rajesh21a,
    title = {Shock Propagation Following an Intense Explosion: Comparison Between Hydrodynamics and Simulations},
    author = {Joy, J. P. and Pathak, S. N.  and  Rajesh, R.},
    journal = {J. Stat. Phys.},
    volume = {182},
    pages = {34},
    numpages = {22},
    year = {2021},
    url = {https://doi.org/10.1007/s10955-021-02715-3}
}

@article{Rajesh21b,
    title = {Shock propagation in the hard sphere gas in two dimensions: comparison between simulations and hydrodynamics},
    author = {Joy, J. P. and Rajesh, R.},
    journal = {J. Stat. Phys.},
    volume = {184},
    pages = {3},
    numpages = {16},
    year = {2021},
    url = {https://doi.org/10.1007/s10955-021-02790-6}
}

@article{de2019,
  title={Diffusion in generalized hydrodynamics and quasiparticle scattering},
  author={De Nardis, J. and Bernard, D. and Doyon, B.},
  journal={SciPost Physics},
  volume={6},
  number={4},
  pages={049},
  year={2019}
}

@article{zaburdaev2011,
  title={Perturbation spreading in many-particle systems: a random walk approach},
  author={Zaburdaev, V. and Denisov, S and H{\"a}nggi, P.},
  journal={Phys. Rev. Lett.},
  volume={106},
  number={18},
  pages={180601},
  year={2011},
  publisher={APS}
}

@article{zhao2006,
  title={Identifying diffusion processes in one-dimensional lattices in thermal equilibrium},
  author={Zhao, H.},
  journal={Phys. Rev. Lett.},
  volume={96},
  number={14},
  pages={140602},
  year={2006},
  publisher={APS}
}

@article{das2014,
  title={Numerical test of hydrodynamic fluctuation theory in the Fermi-Pasta-Ulam chain},
  author={Das, S. G. and Dhar, A. and Saito, K. and Mendl, C. B and Spohn, H.},
  journal={Phys. Rev. E},
  volume={90},
  number={1},
  pages={012124},
  year={2014},
  publisher={APS}
}

@article{gallagher2019,
  title={From Newton to Navier--Stokes, or how to connect fluid mechanics equations from microscopic to macroscopic scales},
  author={Gallagher, I.},
  journal={Bulletin of the American Mathematical Society},
  volume={56},
  number={1},
  pages={65--85},
  year={2019}
}

@article{saito2021,
  title={Microscopic theory of fluctuating hydrodynamics in nonlinear lattices},
  author={Saito, K. and Hongo, M. and Dhar, A. and Sasa, S.},
  journal={Phys. Rev. Lett.},
  volume={127},
  number={1},
  pages={010601},
  year={2021},
  publisher={APS}
}

@article{visscher1974,
  title={Transport processes in solids and linear-response theory},
  author={Visscher, W. M.},
  journal={Phys. Rev. A},
  volume={10},
  number={6},
  pages={2461},
  year={1974},
  publisher={APS}
}

@article{AKR2008,
  title = {Exciting hard spheres},
  author = {Antal, T. and Krapivsky, P. L. and Redner, S.},
  journal = {Phys. Rev. E},
  volume = {78},
  issue = {3},
  pages = {030301},
  numpages = {4},
  year = {2008},
  doi = {10.1103/PhysRevE.78.030301},
  url = {https://link.aps.org/doi/10.1103/PhysRevE.78.030301}
}

@article{rajesh22,
    title = {Blast Waves in Two and Three Dimensions: {E}uler Versus {N}avier-{S}tokes Equations},
    author = {Kumar, A.  and  Rajesh, R.},
    journal = {J. Stat. Phys.},
    volume = {188},
    pages = {12},
    numpages = {14},
    year = {2022},
    url = {https://doi.org/10.1007/s10955-022-02933-3}
}

@article{kumar2025,
  title={Shock Propagation in a Driven hard-sphere Gas: Molecular Dynamics Simulations and Hydrodynamics},
  author={Kumar, A. and Rajesh, R.},
  journal={J. Stat. Phys.},
  volume={192},
  number={9},
  pages={121},
  year={2025},
  publisher={Springer},
  url = {https://doi.org/10.1007/s10955-025-03503-z}
}

@article{chakraborti2021,
  title={Blast in a One-Dimensional Cold Gas: From Newtonian Dynamics to Hydrodynamics.},
  author={Chakraborti, S. and Ganapa, S. and Krapivsky, P. L. and Dhar, A.},
  journal={Phys. Rev. Lett.},
  volume={126},
  number={24},
  pages={244503--244503},
  year={2021}
}

@article{taylor1950formation,
  title={The formation of a blast wave by a very intense explosion I. Theoretical discussion},
  author={Taylor, G. I.},
  journal={Proceedings of the Royal Society of London A},
  volume={201},
  number={1065},
  pages={159--174},
  year={1950}
}

@article{vonneumann1947blast,
  title={The point source solution},
  author={von Neumann, J.},
  journal={Collected Works},
  volume={6},
  pages={219--237},
  year={1963}
}

@ARTICLE{sedov1946,
       author = {{Sedov}, L. I.},
        title = "{Propagation of strong shock waves}",
      journal = {Journal of Applied Mathematics and Mechanics},
         year = 1946,
        month = jan,
       volume = {10},
        pages = {241-250},
       adsurl = {https://ui.adsabs.harvard.edu/abs/1946JApMM..10..241S}
}

@article{ganapa2021blast,
  author = {Ganapa, S. and Chakraborti, S. and Krapivsky, P. L. and Dhar, A.},
  title = {Blast in the one-dimensional cold gas: Comparison of microscopic simulations with hydrodynamic predictions},
  journal = {Physics of Fluids},
  volume = {33},
  number = {8},
  pages = {087113},
  year = {2021},
  doi = {10.1063/5.0058152}
}

@book{landau1987fluid,
  author    = {Landau, L. D. and Lifshitz, E. M.},
  title     = {Fluid Mechanics},
  series    = {Course of Theoretical Physics},
  volume    = {6},
  edition   = {2nd},
  publisher = {Pergamon Press},
  address   = {Oxford},
  year      = {1987}
}

@book{leveque1992numerical,
  author    = {LeVeque, R. J.},
  title     = {Numerical Methods for Conservation Laws},
  series    = {Lectures in Mathematics ETH Z{\"u}rich},
  edition   = {2nd},
  publisher = {Birkh{\"a}user},
  address   = {Basel},
  year      = {1992}
}

@article{Mendl_2016,
   title={Shocks, Rarefaction Waves, and Current Fluctuations for Anharmonic Chains},
   volume={166},
   ISSN={1572-9613},
   url={http://dx.doi.org/10.1007/s10955-016-1626-5},
   DOI={10.1007/s10955-016-1626-5},
   number={3-4},
   journal={J. Stat. Phys.},
   publisher={Springer Science and Business Media LLC},
   author={Mendl, C. B. and Spohn, H.},
   year={2016},
   month=Oct, pages={841–875} }

@book{spohnbook,
author = {Spohn, H.},
title = {Hydrodynamic Scales of Integrable Many-Body Systems},
publisher = {World Scientific},
year = {2024},
doi = {10.1142/13600},
address = {},
edition   = {},
URL = {https://www.worldscientific.com/doi/abs/10.1142/13600},
eprint = {https://www.worldscientific.com/doi/pdf/10.1142/13600}
}

%\nolinenumbers

%##########################################################################

\end{document}